# Crack-Net: Prediction of Crack Propagation in Composites

Hao Xu[1,†], Wei Fan[2,†,*], Ambrose C. Taylor[2], Dongxiao Zhang[3,4,*], Lecheng Ruan[5], Rundong Shi[6]

[1] *BIC-ESAT, ERE, and SKLTCS, College of Engineering, Peking University, Beijing 100871, P. R. China*

[2] *Department of Mechanical Engineering, Imperial College London, South Kensington Campus, London, SW7 2AZ, UK*

[3] *Eastern Institute for Advanced Study, Eastern Institute of Technology, Ningbo, Zhejiang 315200, P. R. China*

[4] *Department of Mathematics and Theories, Peng Cheng Laboratory, Shenzhen 518000, Guangdong, P. R. China*

[5] *Department of Mechanical and Aerospace Engineering, University of California, Los Angeles, Los Angeles, CA, 90095, U.S.A.*

[6] *Meituan Corporate*

\* Corresponding authors. Email address: w.fan19@imerpial.ac.uk (W. Fan); dzhang@eias.ac.cn (D. Zhang).

† These authors contributed equally.

**Abstract**

Computational solid mechanics has become an indispensable approach in engineering, and numerical investigation of fracture in composites is essential as composites are widely used in structural applications. Crack evolution in composites is the bridge to elucidate the relationship between the microstructure and fracture performance, but crack-based finite element methods are computationally expensive and time-consuming, limiting their application in computation-intensive scenarios. Here we propose a deep learning framework called Crack-Net, which incorporates the relationship between crack evolution and stress response to predict the fracture process in composites. Trained on a high-precision fracture development dataset generated using the phase field method, Crack-Net demonstrates a remarkable capability to accurately forecast the long-term evolution of crack growth patterns and the stress-strain curve for a given composite design. The Crack-Net captures the essential principle of crack growth, which enables it to handle more complex microstructures such as binary co-continuous structures. Moreover, transfer learning is adopted to further improve the generalization ability of Crack-Net for composite materials with reinforcements of different strengths. The proposed Crack-Net holds great promise for practical applications in engineering and materials science, in which accurate and efficient fracture prediction is crucial for optimizing material performance and microstructural design.

**Introduction**

Fracture are one of the most common failure modes in engineering applications, and numerical prediction of ruptures based on microstructural information is of great significance. This problem is especially urgent and complex in composites, which are extensively used in aerospace, automotive, watercraft, wind turbines, etc. The modelling of crack evolution using composite representative volume elements (RVEs) plays a key role in bridging between the microstructure and the fracture properties, understanding the fracture mechanisms, and facilitating material design. Crack-based finite element methods (FEM), such as the cohesive zone model[1,2], the damage model via Abaqus Subroutine[3,4], extended FEM[5–7] and crack phase field approach[8–11], have been widely investigated, and remarkable insights into the fracture process have been obtained through these simulations. However, the performance of FE models is highly dependent on fine element sizes and small increment sizes, which are always computationally demanding and time-consuming. The shortcoming of expensive computation is furthermore augmented when batches of simulations are necessary, such as high-throughput computation for screening and optimization of composite compositions.

In recent years, artificial intelligence (AI) models trained on simulation data have attracted increasing attention due to their remarkable capacity for predicting the mechanics of materials. Deep neural networks (DNN) were commonly used in computational mechanics to provide direct predictions from the undeformed microstructural RVE to the desired outputs, such as mechanical properties[12–15], stress-strain curves[16–23], stress fields[24–28], and crack patterns[24,28–30], and good agreement between these predictive models and the results from FE simulations have been shown. Nevertheless, these models often necessitate a substantial volume of training data, ranging from several thousand to hundreds of thousands of data for specific scenarios, and their applicability across diverse scenarios remains limited. The primary reason for this is that such long-term (from undeformed state to the final state) predictions are end-to-end (no processing information) and brute force (purely data driven, no physics are learned). For instance, one 10,000-element simulation case might consume enormous time and generate gigabyte-level data. However, when training the AI model, only the initial microstructural morphology and specific outcomes are extracted, leading to a limited utilization of the available information. Alternatively, spatiotemporal dynamical frameworks have become a new type of surrogating FE models[31–36]. In this approach, the spatial patterns were captured by the DNN solver for partial differential equations (PDEs) under varied initial/boundary conditions, and time marching was achieved by looping the PDE solver. However, such approaches require explicit analytical equations for different problems, which is hard to build in complex composite microstructures. Furthermore, large training datasets are needed for a robust PDE solver.

To handle the above-mentioned challenges, we present Crack-Net in this work, which is able to simultaneously predict crack growth patterns and stress response. The key innovation is that the relationship between the composite microstructure and stress response is bridged by the crack evolution process, and this inherent capture of the physical nature of the process endows the framework with the following features:

- The underlying relationship between stress response and crack growth path is incorporated in the architecture of Crack-Net, making the predictions much more reliable.
- The demand for the training datasets is reduced by up to two-orders of magnitude, with only 554 simulation cases from particle reinforced composite (PRC) materials with varied volume fractions being used. In comparison, AI-based end-to-end predictions of merely stress-strain curve in composites typically requires thousands to tens of thousands of cases[16–20].
- The Crack-Net model presents remarkable generalization ability among microstructural patterns with distinguishing features. It not only performs perfectly on random PRCs but also shows impressive out-of-sample predictive ability on interpenetrating phase composites (IPCs), which presents distinctly different microstructural pattern from PRCs (Fig. 3d).
- The modulus of the reinforcing component has a significant effect on the crack propagation path and stress response, and research dealing with different component properties is scarce. Crack-Net integrates transfer learning technology to augment its flexibility and generalization ability. By employing only 120 simulations with new component properties, Crack-Net's ability is extended to new scenarios.

- Compared with crack phase field simulations used for the training dataset, Crack-Net reduces the running time by approximately 95%, and only takes a few seconds for a short-term prediction (single step) and 1-2 minutes for a 150-step long-term prediction.

We believe that Crack-Net will offer engineers and researchers the expedient capacity to assess crack evolution and comprehend pivotal facets of the fracture process. This functionality will guide critical tasks, such as composite design, the study of mechanisms and failure analysis, thus enhancing the efficiency and effectiveness of engineering practices.

**Design philosophy and realization of Crack-Net**

As mentioned above, understanding crack evolution in composites is the key step to bridge the between the microstructure and fracture performance. Although DNN-based crack path predictions in fiber reinforced composites (FRCs) have been studied[24,28,29], the models are not suitable for complicated crack patterns, such as when rupture of the secondary phase is involved. Moreover, the crack evolution predictions did not contribute to the prediction of mechanical properties, such as stress-strain curves.

In this work, a spatiotemporal dynamical workflow is designed to predict the mechanical performance and crack growth path of composites under tensile displacement control. In Fig. 1a, the continuous fracture process of PRC RVEs is evenly divided into discrete frames, and each frame is characterized by the spatial crack pattern and the stress applied to the whole domain. Accordingly, a hierarchical approach was proposed (Fig. 1b), i.e., short-term spatial predictions and long-term time marching. In the short-term hierarchy, the spatial crack pattern and overall stress at step $n$ are predicted based on the information from step $n$-1 via the end-to-end Crack-Net model rather than solving PDEs, and at the long-term loop, the Crack-Net model is called iteratively to generate the whole fracture process of composite RVEs from the undeformed state to the rupture state. The architecture of end-to-end Crack-Net (short-term predictions) is displayed in Fig. 1c.

The training dataset comprising 554 cases was obtained through phase field simulations[9,37] on a 100×100-element PRC domain (Extended Data Fig. 1a), and the cracks are described by a continuous scalar, crack phase field $d \in [0, 1]$, where $d=0$ indicates the intact material and $d=1$ indicates a fully broken state. The incremental strain $\Delta\varepsilon$ of 0.01% was applied to partition the fracture process, and 83,480 step pairs (the current step and the next step) were extracted from the 554 simulation cases. The step pairs were then divided into three sets, allocating 80% for training (66,784 steps), 10% for validation (8,348 steps), and another 10% for testing (8,348 steps) purposes. The training process involved 3,000 epochs with a batch size of 100, and to prevent overfitting, an early stopping technique was used. Importantly, the dataset was randomly split based on the step in the slice prediction. In the dataset, the spatial distribution of matrix phase A and reinforcing phase B is characterized by material parameters in the matrix ($E_{A0}$=5000, $UTS_{A0}$=35, $E_{B0}$=500, $UTS_{B0}$=10.8). The resulting morphology patterns in the training process are illustrated in Fig. S23, which shows that the minutiae of the prediction become more apparent as the training epoch increases.

Apparently, precise incremental predictions from step to step are essential to the long-term process of Crack-Net. Here, the strain increment between steps, $\Delta\varepsilon$, is 0.01%. The differences in the crack phase field between steps are generally subtle, particularly during the initial and final stages of fracture development, where the difference of the crack phase field usually ranges from $10^{-2}$ to $10^{-5}$. Here, instead of the crack phase field matrix, **d**, a transformed incremental value matrix $\Delta$**d'** is predicted. For each element, as the value of $d$ is between 0 and 1 and keeps monotonically increasing from step k to step k+1, $\Delta d_k = d_{k+1} - d_k$ is usually a tiny number between 0 and 1. Therefore, log transformation is adopted to amplify minor differences in the crack phase field, and the transformed incremental crack phase field, $\Delta d'_k$, is defined as follows:

$$\Delta d'_k = 10 + \log_{10}(\max(10^{-10}, d_{k+1} - d_k)), (0 \leq \Delta d'_k \leq 10) \qquad (1)$$

This approach manages to capture the fine-grained details of crack propagation, enabling Crack-Net to make more precise and reliable predictions in scenarios where the development of the crack phase field is subtle. As a consequence, the input of Crack-Net involves the initial elastic modulus matrix (**E₀**) and ultimate tensile strength matrix (**UTS**), crack phase field matrix (**d_k**), and the applied strain ($\varepsilon_k$) and stress ($\sigma_k$) in step k. The output is the applied stress in the next step ($\sigma_{k+1}$) and the transformed increment of crack phase field $\Delta d'_k$. Additional details about the construction of the Crack-Net are provided in the Materials and Methods section.

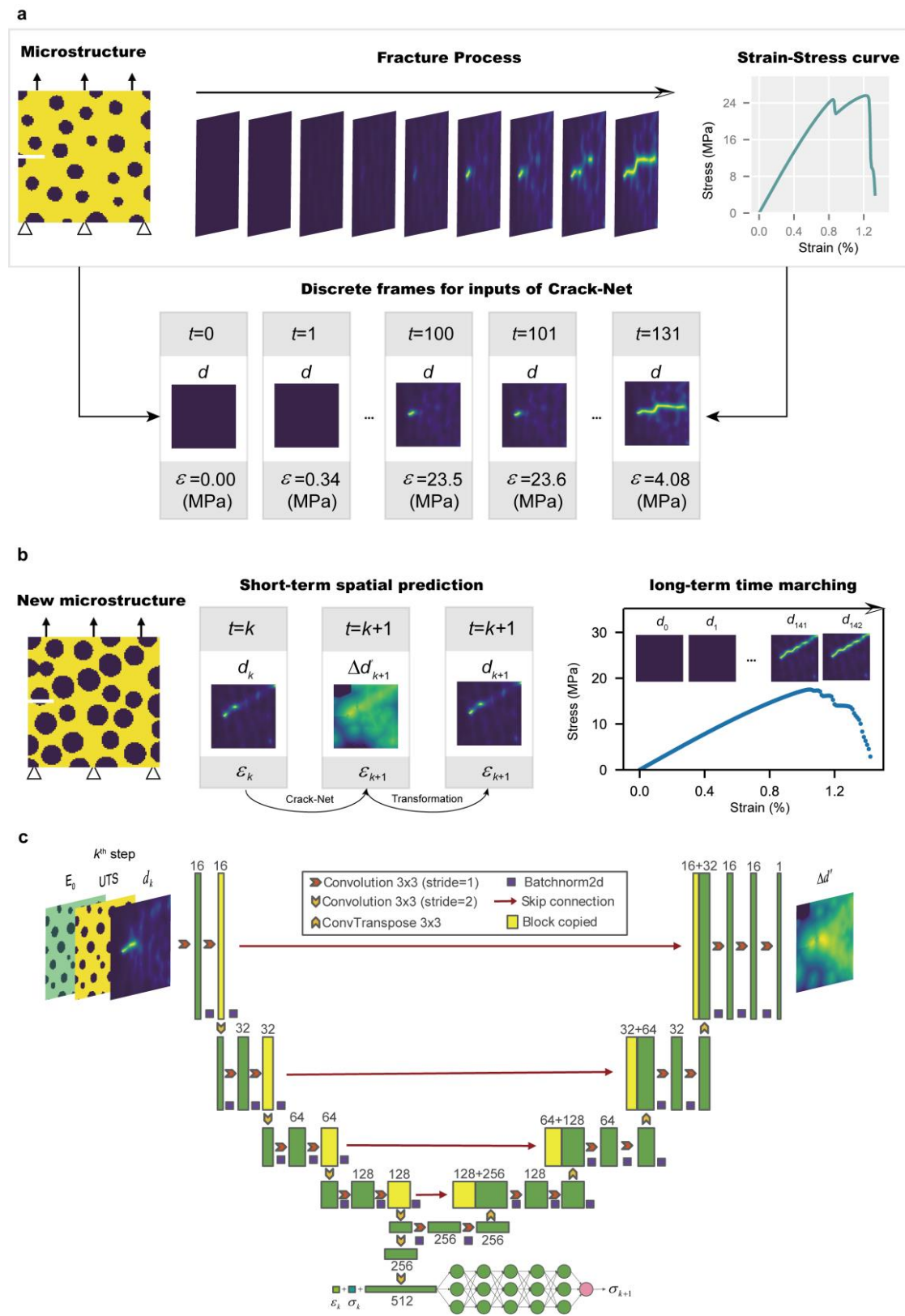

**Fig. 1. The framework of the Crack-Net. (a)** The input information of Crack-Net. The fracture process and stress-strain curve are evenly divided into discrete frames, and each frame is characterized by the spatial crack pattern and stress applied to the whole domain. **(b)** The prediction procedure of Crack-Net, including short-term spatial prediction and long-term time marching. In short-term prediction, the transformed increment of crack phase field $\Delta d'_{k+1}$ and overall stress $\varepsilon_{k+1}$ at step $k$ are predicted based on the information from step $k$-1, and the crack pattern $d_{k+1}$ at step $k$ is transformed from the predicted increment. **(c)** The architecture of Crack-Net, which utilizes the composite design, fracture phase field, strain, and stress in the current step to predict the transformed increment of the fracture phase field and stress in the next step, which are coupled by the same latent vector.

**Results**

**Short-term incremental prediction**

In this section, we focus on evaluating the short-term prediction performance of Crack-Net, which constitutes a fundamental aspect that significantly influences the overall prediction capability of the model. The short-term prediction performance of the trained Crack-Net is depicted in Fig. 2. Given its capability to simultaneously predict stress response and crack evolution, the performance of both predictions is examined. As shown in Fig. 2a, the predicted stress demonstrates an excellent match with the observations from the testing dataset, achieving a remarkable $R^2$ value of 0.9993 and a small mean squared error (MSE) of 0.0378. The agreement between predictions and observations serves as compelling evidence that confirms the predictive ability of Crack-Net to precisely capture the development of stress response during crack propagation in the short-term hierarchy. Considering that the crack phase field is a two-dimensional field, we use the overall relative error to measure the accuracy of the prediction, which is defined as:

$$e_d = \sqrt{\frac{\sum_i \sum_j (\hat{d}_{i,j} - d_{i,j})^2}{\sum_i \sum_j (d_{i,j})^2}}, \quad (2)$$

$$e_{\Delta d'} = \sqrt{\frac{\sum_i \sum_j (\Delta \hat{d}'_{i,j} - \Delta d'_{i,j})^2}{\sum_i \sum_j (\Delta d'_{i,j})^2}}, \quad (3)$$

where $e_d$ and $e_{\Delta d'}$ are the relative error of the predicted phase field $\hat{d}_{i,j}$ and the transformed increment $\Delta \hat{d}'_{i,j}$, respectively; and *i* and *j* are the index of row and column in the field, respectively. Notably, the output of Crack-Net is the transformed increment of the phase field, and the predicted phase field for the next step is obtained by calculating the absolute increment and adding it to the phase field of the current step, as detailed in the Materials and Methods section. To assess the accuracy of Crack-Net's slice prediction for the crack phase field, we analyze the distribution of relative errors on the testing dataset, as illustrated in Fig. 2b and Fig. 2c. It is discovered that the relative error is less than 1% for over 50% of the testing data, which demonstrates the high accuracy and efficacy of Crack-Net in incremental prediction.

To intuitively demonstrate Crack-Net's slicing prediction capability, we generated new composite designs that are not in the training dataset for out-of-sample slice prediction, and the prediction for different stages of the crack propagation are depicted in Fig. 2. As illustrated in Fig. 2d, the prediction of crack initiation is presented. At this stage, the material remains unbroken, resulting in a minimal value of the crack phase field. However, the subtle accumulation of increments during this phase plays a critical role in the final crack development. Crack-Net exhibits remarkable accuracy in predicting the distribution of these subtle increments of the crack phase field, which forms a solid foundation for long-term fracture propagation predictions. Fig. 2e illustrates the prediction during the unstable crack growth stage. Significant changes in the crack phase field occur between consecutive steps during this stage, which are accurately captured by Crack-Net. Furthermore, the prediction in the last stage of crack propagation is shown in Fig. 2f, in which the crack has fully grown, and the increment of the crack phase field becomes subtle and clustered near the crack. It is found that Crack-Net demonstrates precise predictions for this behavior. Notably, the particle size and proportion of the reinforcement phase in Fig. 2f is obviously different from those in Fig. 2d and Fig. 2e, highlighting Crack-Net's generalization ability for diverse composite designs. Meanwhile, it is observed that even during the unstable crack growth stage, the changes in the crack phase field between steps are not particularly substantial, which implies that directly predicting the crack phase field may overlook crucial details of these incremental changes. Conversely, the transformed increment exhibits notable variations at each stage of crack propagation, rendering it advantageous for network learning and prediction. The emphasis on predicting the transformed increment allows for a more comprehensive and accurate representation of crack propagation dynamics, thus enhancing the effectiveness of Crack-Net's predictions.

Furthermore, we challenged Crack-Net with a harder task to predict crack propagation in IPC RVEs with complex co-continuous structures (Fig. 2g), which features a completely different distribution of the reinforcement phase compared to particle reinforced composites. In binary co-continuous structures, the matrix phase and the reinforcement phase coexist in a continuous and interconnected manner, and form a network-like microstructure. The co-continuous nature results in a complicated crack growth process and thus increases the predictive difficulty. Nevertheless, as observed in Fig. 2g, Crack-Net is still able to provide accurate predictions, successfully capturing the distribution of crack increments. It is essential to emphasize that the training data of Crack-Net only consist of PRC specimens without encountering co-continuous ones. This potentially implies that Crack-Net has learned the effect of crack propagation on stress response, which enables it to predict fractures in composites with diverse structural configurations.

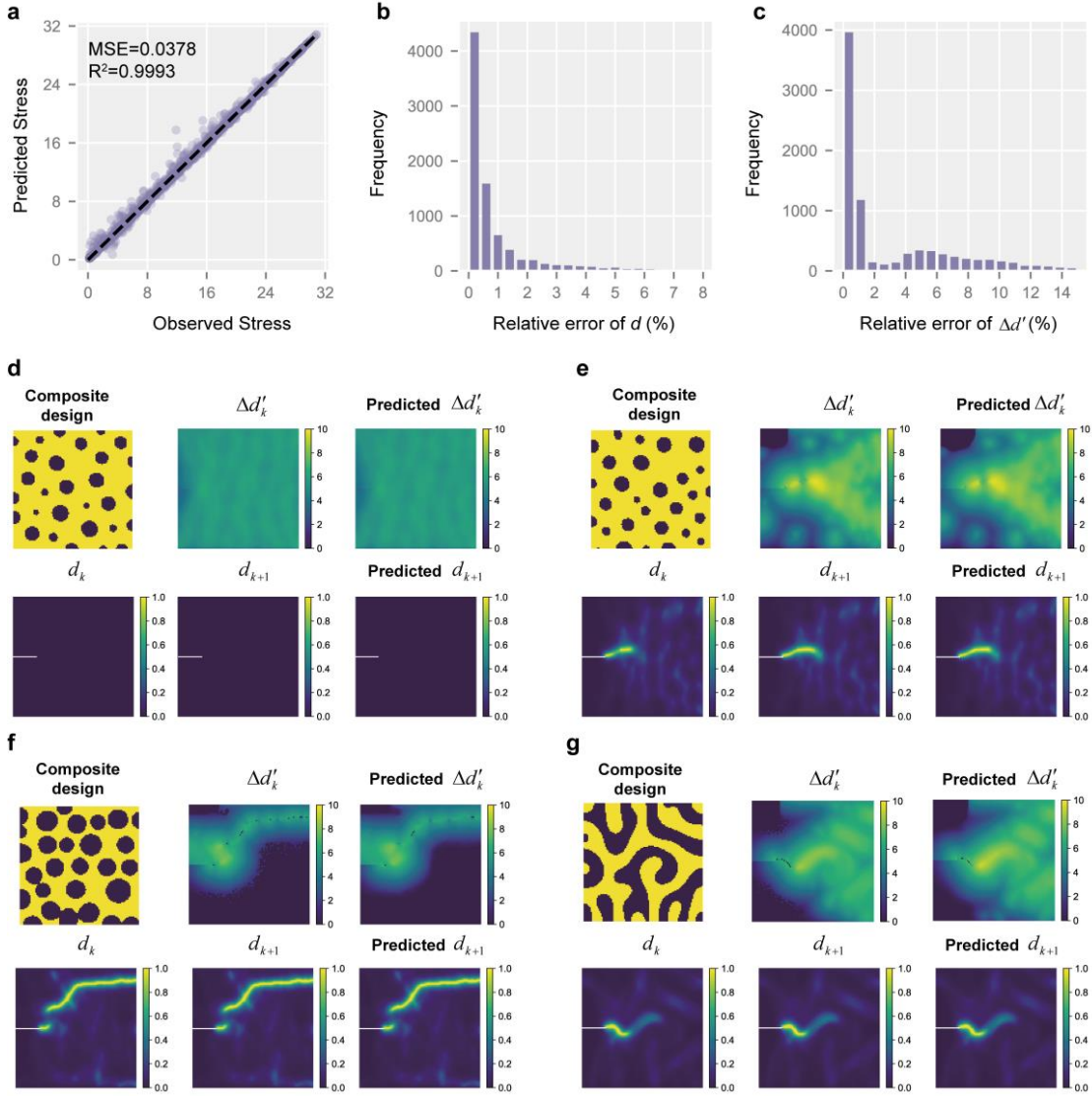

**Fig. 2. The single-step slice prediction of Crack-Net. (a)** The predicted and actual (FE) stress in the testing data. **(b)** The distribution of the relative error of the crack phase field. **(c)** The distribution of the relative error of the transformed increments of the crack phase field. **(d)** The out-of-sample prediction in the crack initialization stage where the composite is unbroken. **(e)** The out-of-sample prediction in the stage of unstable crack growth. **(f)** The out-of-sample prediction in last stage of crack propagation. **(g)** The prediction for binary co-continuous structures. Here, $d_k$ and $d_{k+1}$ are the observed fracture phase field in the current and next step, respectively, and $\Delta d'_k$ is the observed transformed increment. The white line in the phase field represents the initial condition.

**Long-term loop prediction**

In this section, with a given composite microstructure, Crack-Net was employed to predict the fracture process from undeformed state to rupture. The initial crack phase field is termed as (**d₀**, $\sigma_0$, $\varepsilon_0$), where **d₀** is the initial crack phase field matrix filled with 0, while $\sigma_0$ and $\varepsilon_0$ are 0. The initial state was fed into Crack-Net to predict the subsequent step's state, which is then used as input to predict the preceding steps' states of crack. The formula can be written as follows:

$$(\hat{d}_k, \hat{\sigma}_k, \varepsilon_k) \xrightarrow{\text{Crack–Net}} (\hat{d}_{k+1}, \hat{\sigma}_{k+1}, \varepsilon_k + \Delta\varepsilon), \tag{4}$$

$$\hat{d}_{k+1} = \hat{d}_k + 10^{\Delta \hat{d}'_k - 10}. \tag{5}$$

The detailed procedure of the long-term process is provided in the Materials and Methods section and Extended Data Table 1. Apart from pre-determined composite information, Crack-Net can generate predictions without requiring additional conditions or historic states, which enhances its practical applicability. Its long-term prediction capability was assessed on out-of-sample composite RVEs, the results of which are presented in Fig. 3. Several representative steps, i.e., stress peak, crack growth, and rupture point, were extracted from the observed and predicted cases, respectively. Their stress-strain information and crack evolution pattern were also compared. Fig. 3a displays a typical stress-strain curve, in which stress initially increases with strain and rapidly decreases when the crack begins to grow. It can be seen that the predicted stress-strain curve by Crack-Net agrees well with the reference from FE simulations. Meanwhile, it is discovered that the relationship between the stress-strain curve and the crack propagation process is learned well.

Fig. 3b provides a more complex example with two maximum points in the stress-strain curve (R1 and R3 in Fig. 3b), which is challenging for traditional stress-strain curve prediction. In contrast, Crack-Net overcomes this difficulty and provides predictions that closely agree with the FE results. It accurately predicts the two maximum stress points in the stress-strain curve (P1 and P3 in Fig. 3b) and the corresponding crack phase field. In this case, where the stress-strain curve exhibits a unique pattern, it is not sufficient to understand the fracture behavior of materials solely from their stress-strain curves. Fortunately, the predicted crack phase field provides intuitive insights into the crack propagation process by capturing the development of the crack phase field. From Fig. 3b, it is evident that the composite material undergoes fracture from P1 to P2, but the crack only partially propagates and does not fully extend. Consequently, the composites can withstand additional stress during the P2 to P3 stages until the crack eventually grows, which explains the temporary rise of the stress at this stage. This capability of Crack-Net to unveil the intricate details of crack propagation through the predicted crack phase field contributes to a deeper understanding of the fracture process, providing valuable information for further analysis and engineering applications.

In Fig. 3c, notable changes occur in the particle proportion and size for the reinforcing phase in the composites. Despite slight deviations in the predicted stress-strain curve, the characteristic steps are well captured by Crack-Net, especially the brief stress maintenance between P2 and P3 in Fig. 3c. The corresponding crack phase field predicted by Crack-Net remains highly accurate, although the crack is predicted to begin slightly earlier than the reference. Meanwhile, we observed that in this case, the cracks did not propagate continuously (R2 and P2 in Fig. 3c). Instead, they generated independent fracture points at multiple locations which eventually expanded and connected as the strain ratio increased. Remarkably, Crack-Net accurately captured this intricate fracture mode, demonstrating its ability to effectively predict and understand complex crack dynamics, including non-continuous crack propagation. In order to make the result more convincing, further out-of-sample prediction cases and discussions are provided in Supplementary Information S1.1 and Fig. S1-S15.

Finally, we also employed a binary co-continuous structure to further test the generalization ability of Crack-Net, as shown in Fig. 3d. Faced with this type of complex composite structure which is not included in the training dataset, Crack-Net still demonstrates good long-term prediction ability. Although the predicted stress-strain curve in a co-continuous specimen presents larger deviations compared to the one in particle reinforced composites, the most important strength, strain at rupture and toughness are well predicted, and the crack growth process is accurate. More out-of-sample prediction cases and discussions for binary co-continuous structures are provided in Supplementary Information S1.1 and Fig. S17-S19. From the above experiments, Crack-Net is shown to possess satisfactory long-term prediction ability, enabling rapid prediction of crack propagation process in composite materials with diversified structures.

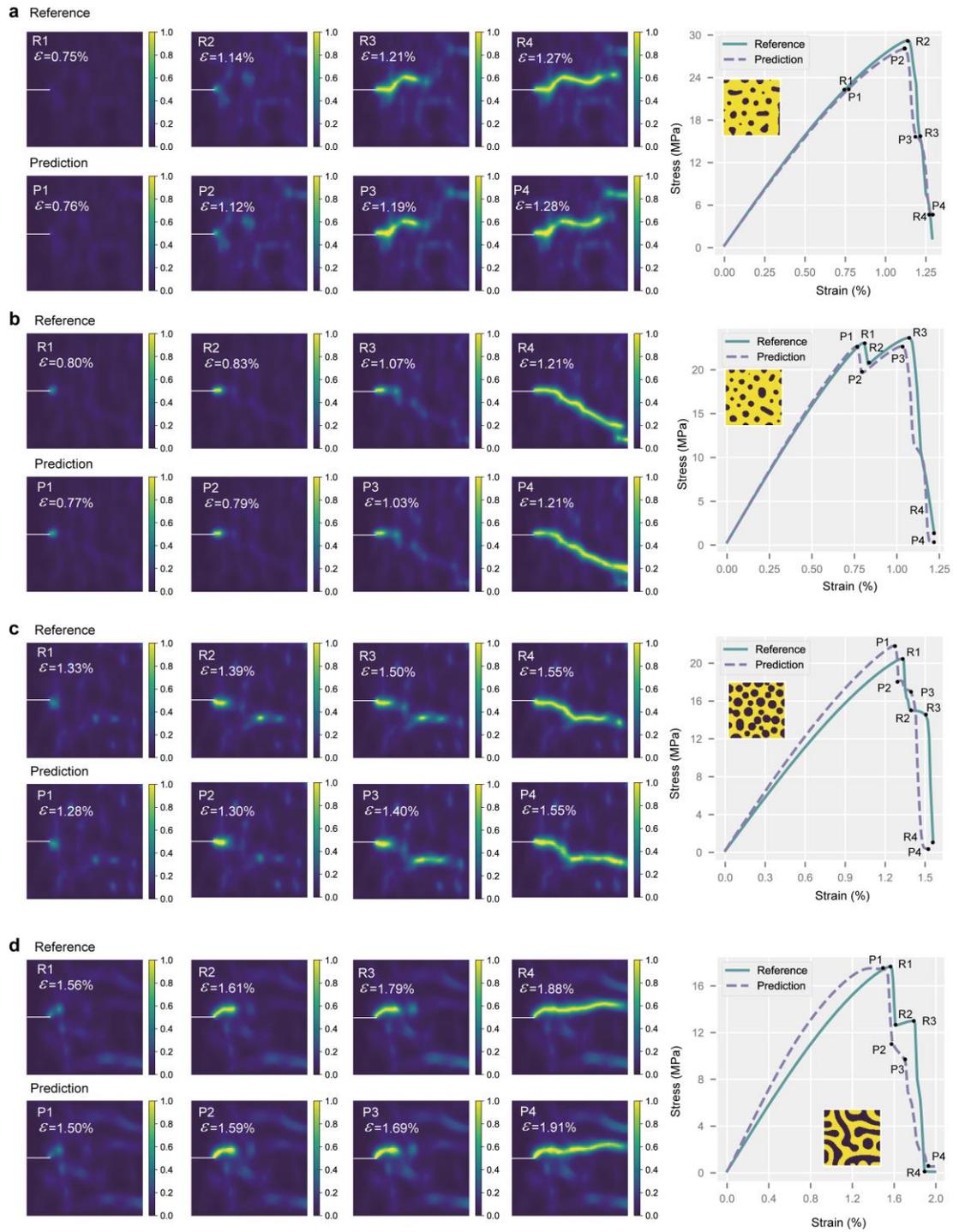

**Fig. 3. The long-term prediction of Crack-Net. (a)-(c)** The observed (R1 to R4) and predicted phase field (P1 to P4) at characteristic steps in the stress-strain curves (left), and the observed and predicted stress-strain curve (right) for the particle-reinforcement composite. **(d)** The observed and predicted phase field at characteristic steps in the stress-strain curves (left), and the observed and predicted stress-strain curve (right) for the binary co-continuous composite. The composite microstructure design is provided beside the stress-strain curve, where the yellow region is the matrix phase, and the black part region is the reinforcement phase. $\varepsilon$ is the strain ratio. The white line in the phase field represents the initial condition.

**Transfer learning for composites with different material properties**

While Crack-Net exhibits commendable predictive ability for diverse composite designs, the material properties for the reinforcement phase and matrix in the above cases are fixed. This implies that Crack-Net's current capabilities are limited to predicting crack propagation exclusively with specific material properties, which will constrain its practical applicability. However, directly predicting composites with different material properties necessitates a substantial amount of training data, which is challenging to acquire due to the considerable time cost associated with numerical simulations. Obtaining a sufficiently diverse and comprehensive dataset that encompasses the vast range of possible material properties is impractical within realistic time constraints. This limitation poses a significant hurdle in achieving accurate predictions for composite materials with varying material properties using traditional training methods. Therefore, transfer learning presents a promising solution to address this challenge, as it enables Crack-Net to leverage knowledge learned from one set of material properties and adapt it to predict crack propagation dynamics in composites with different material properties. With a smaller amount of data in transfer learning, Crack-Net can effectively generalize and enhance its predictive capabilities for a broader spectrum of material properties without the need for an exhaustive dataset covering all conceivable material properties. As depicted in Fig. 4a, the current Crack-Net is trained in the source domain using a substantial dataset, demonstrating its high accuracy in predicting the stress-strain curve and crack phase field. In the context of transfer learning, the parameters of this well-trained model are transferred to a new prediction model. This new model, initialized with the knowledge from the source domain, only requires a small number of training data in the target domain for fine-tuning. Through this transfer process, the model becomes capable of predicting the stress-strain curve and crack propagation process in composites with different material properties. More information about the implementation of transfer learning in Crack-Net is provided in the Materials and Methods section.

From a fracture mechanism perspective, the propagation of cracks in composite materials is notably influenced by the relative strength of the matrix and reinforcement phase (Fig. S22). Generally, in particle-reinforced composite materials, when the reinforcement phase possesses higher strength, cracks tend to develop in the matrix (Fig. 4a). Conversely, when the reinforcement phase has lower strength, cracks are more prone to propagate through the reinforcement phase. For this reason, for the target domain of transfer learning, we focus on two typical situations where the strength of the reinforcement phase is large and small, respectively. These two situations exhibit divergent crack propagation patterns due to the distinct material property settings.

Fig. 4b demonstrates the transfer learning performance of Crack-Net for weak reinforcement strength. In this scenario, the strength of the reinforcement phase is reduced to $UTS_{B0}$=4.8 and $E_{B0}$=100, signifying a significantly lower strength compared to the matrix. In the transfer learning, 120 cases with different composite designs are adopted as the dataset to fine-tune the model. The detailed settings for the training process are provided in the Materials and Methods section. We conducted tests on several newly-generated microstructures to assess the out-of-sample prediction performance of the transfer learning model, and compared the predictions to the source models without transfer learning. As illustrated in Fig. 4b, the results demonstrated a significant improvement in prediction accuracy with transfer learning.

Similarly, transfer learning was applied to composites with strong reinforcement phases ($UTS_{B0}$=112 and $E_{B0}$=50,000). The training dataset also consisted of 120 cases, and the training setting remained the same. The outcomes demonstrated the effectiveness of transfer learning in enhancing the model's predictive generalization ability, as well (Fig. 4c). In contrast, the source model's predictions were entirely incorrect, indicating a lack of predictive ability. However, as shown in Fig. S22, it is proven that the model maintains certain predictive ability when the modulus of the reinforcing component is close to the source domain. This finding implies that we do not need to perform transfer learning for numerous material properties. Instead, transfer learning can be applied to several typical situations, allowing the new model to possess good predictive ability for composites with reinforcement strength within a certain range. More discussions are provided in Supplementary Information S1.2.

Meanwhile, the experiments with different data volumes for transfer learning revealed that the augmentation of the dataset for transfer learning directly correlates with an escalation in the predictive accuracy of the model, and the model performs well faced with even 50 cases, which is detailed in Supplementary Information S1.2 and Fig. S20.

Meanwhile, in order to better demonstrate the superiority of the transfer learning, we conducted an additional experiment by comparing the performance of Crack-Net directly trained by the same dataset utilized in the transfer learning. The results are shown in Fig. S21, which indicates that the directly-trained model without transfer learning failed to the predict the correct crack phase field (Supplementary S1.2). This confirms that the learned underlying mechanism of Crack-Net in the source domain can facilitate the training of the model in the target domain, which makes the transfer learning effective in this task.

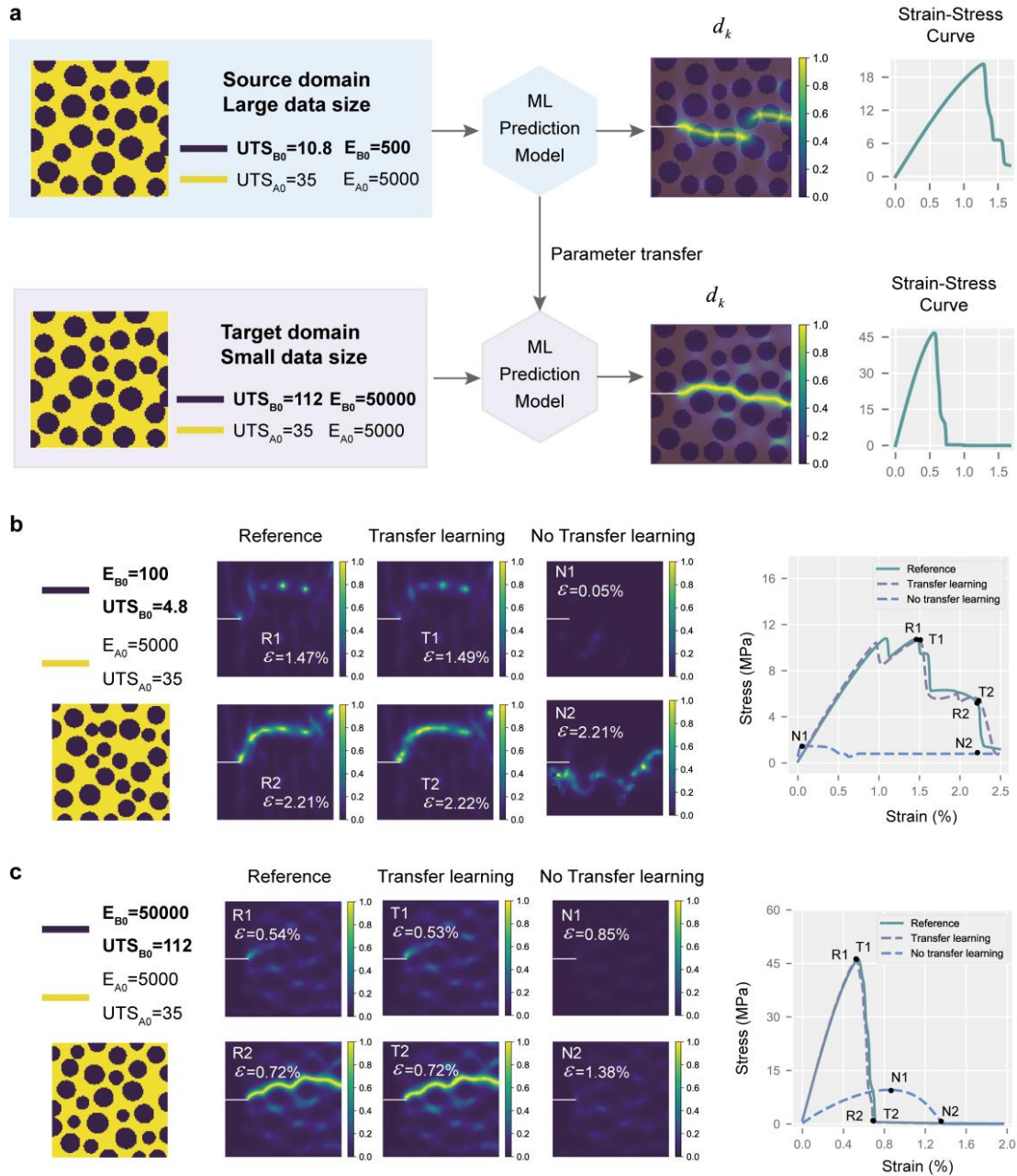

**Fig. 4. The illustration and outcomes for the transfer learning of Crack-Net. (a)** The sketch map for the transfer learning. **(b)** The outcomes of the transfer learning for the composites with a weak reinforcement phase, where the material property of the reinforcement phase is changed. **(c)** The outcomes of the transfer learning for the composites with a strong reinforcement phase, where the material property of the reinforcement phase is changed. $\varepsilon$ is the strain. The white line in the phase field represents the initial condition.

**Discussion**

In this work, a deep-learning framework called Crack-Net is proposed to accurately predict the fracture process in composite materials. The hierarchical architecture of Crack-Net makes it able to learn the underlying relationship between crack evolution and stress response, which provides enhanced predictive accuracy and physical consistency. The model's long-term prediction capability enables it to simultaneously predict the crack propagation process and stress-strain curve for a given composite microstructure. Moreover, the incorporation of the transfer learning technique reduces the demand for the training dataset by 80% when a new scenario appears, for example when the modulus of the reinforcing phase varies significantly.

The practical significance of Crack-Net lies in its transformative impact on the numerical study of the fracture of composites. Benefiting from its accurate and reliable prediction and small demand for training dataset, Crack-Net opens new possibilities for optimizing composite design, screening component compositions, and accelerating high-throughput computation applications. In summary, Crack-Net represents a significant advancement in the study of the fracture of composite materials. Its incorporation of high-precision numerical simulation, deep-learning crack phase field prediction, and transfer learning contributes to the enhancement of predictive accuracy and adaptability. Future works will concentrate on the embedding of explicit physical rules to further improve the model performance, the realization in high-throughput computation scenarios, and the extension of the current quasi-static fracture prediction to more complicated problems when plasticity, dynamic fracture or fatigue, are considered. Despite this, the proposed Crack-Net not only offers valuable insights into the intricate mechanics of crack propagation, but also holds great promise for practical applications in engineering and materials science, where accurate and efficient fracture prediction is crucial for optimizing material performance and design.

**Materials and Methods**

**Morphology development from phase separation simulation**

An A-B binary composite system resulting from phase separation of miscible polymer blends presents good interfacial adhesion between the components, which enhances the overall mechanical performance. The generating morphologies are usually divided into two types, i.e., particle reinforced composites (PRCs) and interpenetrating phase composites (IPCs), depending on the initial compositions, $c_{A0}$ and $c_{B0}$, where $c_{A0} + c_{B0} = 1$. The morphology evolution is governed by the Cahn-Hilliard equation[38]:

$$\frac{\partial c_B}{\partial t} = M\Delta[\frac{df(c_B)}{dc_B} - \gamma \Delta c_B], \tag{6}$$

where $c$ is the concentration of B component; $t$ is time; $M$ is the mobility coefficient; $\gamma$ is a constant regarding interphase thickness; and $f(c)$ is a free energy density function.

In the current study, a 100 μm * 100 μm domain was meshed into identical square elements with element size $h_{FEM}$, where $h_{FEM}$ = 1 μm for the numerical simulation. An open-source FEniCS algorithm[39,40] was employed to generate the morphology, and periodic boundary conditions were adopted to the domain. Time step $dt$ = 5e$^{-6}$ s, mobility $M$ = 1 m$^2$/s, $\gamma$ = 0.01 m$^2$/s, and $f(c) = 100 \cdot c^2 \cdot (1-c)^2$. An initial $c_B$ value was randomly assigned with normal distribution $N(c_{B0}, 0.01)$ to each element at $t = 0$. By varying $c_{B0}$ from 0.25 to 0.5, the morphology varies from PRC to IPC, as shown in Extended Data Fig. 1c. The volume fractions of A-rich phase, $V_{f,A}$, and B-rich phase, $V_{f,B}$, were controlled by a cutoff value of concentration, and different cutoff values were selected for different $c_{B0}$-series. The detailed information is provided in Table S1.

**Fracture simulations and dataset generations**

The resulting specimens were then subjected to fracture simulations, by taking the A-rich phase as matrix and the B-rich phase as modifier. An edge precut of 20 μm was introduced to the matrix. Since the morphology was randomly generated, specimens in which the crack tip was located in the B-rich phase were not used. The specimens were fixed at the bottom boundary, and stretched at the top boundary with a displacement controlled increment size $10^{-3}$ μm (Extended Data Figure 1b).

The phase field method by Gergely Monlar[37] was adopted to simulate the crack evolution process in the domain, where a crack was described by a continuous scalar $d \in [0, 1]$, with $d = 0$ indicating the pristine material and $d = 1$ indicating the ruptured state. Both the A-rich phase and the B-rich phase were considered as elastic solids, and their interfaces were assumed to be perfectly bonded. The fracture behaviour of materials is described by tensile modulus $E$, fracture surface energy density $g_c = 10$ J/m$^2$, and characteristic width, $l_c = 2h_{FEM} = 2$ μm. Modulus of matrix were fixed, $E_A = 5$ GPa, and modulus of modifiers was changed accordingly as displayed in Table S1 and S2. For each simulation, the processing data, including overall applied stress and crack phase field map, was extracted at the output frequency = 10, or, incremental strain = 0.01%. Finally, the initial morphology information of the composite microstructure, the component model parameters, and the processing information such as overall stress-strain curve and crack phase field, consist of the training dataset.

**Crack-Net**

The proposed Crack-Net adopts a specialized architecture inspired by the U-Net, featuring an encoder-decoder design. The U-Net, originally developed for medical image segmentation, has gained widespread popularity due to its remarkable performance and versatility in various applications[41,42]. In the U-Net architecture, the encoder part utilizes a series of convolutional layers to extract hierarchical features from the input data, gradually reducing its spatial dimensions while capturing intricate patterns. The decoder part then employs up-convolutional layers to expand the extracted features and reconstruct the output with the same spatial dimensions as the input. One of the key advantages of U-Net lies in its ability to preserve spatial information during the encoding and decoding process. This characteristic ensures that the fine details and localized features of the input data are retained in the output, making it particularly effective for certain tasks, such as image segmentation and object detection. The skip connections between corresponding encoder and decoder layers further enhance the model's capability to capture both low-level and high-level features, facilitating precise predictions.

In the Crack-Net, the size of the input is 3×100×100, including the initial elastic modulus ($E_0$), ultimate tensile strength (UTS), and the current crack phase field $d_k$, which are all fields with the size of 100×100. The size of output is the transformed increment of the crack phase field $\Delta d_k'$ with the size of 1×100×100. The network architecture is presented in Fig. 1b. The encoder consists of four down-sampling blocks, where two convolutional layers are employed to extract the features. The first convolutional layer has the input channel of *in_ch* and the output channel of *out_ch*, while the second convolutional layer has the input channel of *out_ch* and the output channel of *out_ch*. The kernel size, stride and padding for both convolutional layers is 3, 1, and 1, respectively. In each down-sample block, the output of these convolutional layers is then passed into a convolutional layer with the input channel of *out_ch*, the output channel of *out_ch*, the kernel size of 3, the stride of 2, and the padding of 1. The output of this layer is sent to the corresponding up-sampling block through the skip connections (Fig. 1b). The *in_ch* for the four down-sampling blocks is 3, 16, 32, and 64, respectively. The *out_ch* for the four down-sampling blocks is 16, 32, 64, and 128, respectively. The decoder consists of four up-sampling blocks, where two convolutional layers are employed to further extract the features and one up-convolutional layer to expand the extracted features. The first convolutional layer has the input channel of *in_ch* and the output channel of *out_ch*×2, while the second convolutional layer has the input channel of *out_ch*×2 and the output channel of *out_ch*×2. The kernel size, stride and padding for both convolutional layers is 3, 1, and 1, respectively. For the up-convolutional layer, the input channel is *out_ch*×2, the output channel is *out_ch*, the kernel size is 3, the stride is 2, and the padding is 1. As shown in Fig. 1b, the input of the up-sampling block is the combination of the output feature from the previous block and the feature sent from the skip connection. Therefore, the *in_ch* for the four up-sampling blocks is 64, 128, 64, and 32, respectively. The *out_ch* for the four up-sampling blocks is 128, 64, 32, and 16, respectively. The output of the decoder is then sent to the output block, where three convolutional layers are utilized to reconstruct the output. The input channels of these convolutional layers are 32, 16, and 16, and the output channels are 16, 16, and 1, respectively. The kernel size, stride and padding are 3, 1, and 1, respectively. There is a batch normalization layer and a rectified linear unit (ReLU) layer between all convolutional layers.

The architecture described above is similar to the conventional U-Net structure, which aids in reconstructing the transformation increment of the predicted crack phase field. Meanwhile, our Crack-Net goes beyond this initial reconstruction and implements further processing on the latent vector extracted by the encoder. This crucial step enables the model to achieve simultaneous prediction of stress and crack phase field, making it specialized for precise crack propagation prediction in composite materials. As shown in Fig. 1b, the extracted latent vector is sent into an additional feature extraction block to further extract the information. In the block, there are four convolutional layers, where the input channels of these convolutional layers are 128, 256, 256, and 512, the output channels are 256, 256, 512, and 512, the kernel size is 3, the stride is 2, and the padding is 1. Consequently, the output of the feature extraction block is a 512-dimensional vector, containing high-dimensional latent information concerning the input composite design and the current crack phase field. To further predict the stress in the next step, this 512-dimensional vector is combined with the strain and stress from the current step and fed into a fully-connected artificial neural network (ANN). The ANN consists of an input layer with 514 neurons, two hidden layers with 128 neurons and 32 neurons, respectively, and an output layer with 1 neuron. This sophisticated process enables Crack-Net to accurately predict the stress response in the next step. It is important to emphasize that the predicted stress and crack phase field for the next step are originated from the same latent vector extracted from the encoder part of Crack-Net. This design enables the network to capture the underlying relationship between crack propagation and strain development, enhancing the accuracy and physical coherence of the predictions.

**Long-term prediction by Crack-Net**

Although the Crack-Net is trained by slice data, the relationship between the input and output makes it able to make long-term prediction by iteratively predicting the next step from the previous prediction. In this section, the procedure of long-term prediction of Crack-Net is detailed. As shown in Extended Data Table 1, the long-term prediction only requires the initial situation and the composite design. Without any prior information, Crack-Net is directly employed to predict the crack propagation starting from the initial step, where the fracture has not yet initiated. The initial fracture phase field is termed as $(d_0, \sigma_0, \varepsilon_0)$ where $d_0$ is the initial fracture phase field filled with 0, and $\sigma_0$ and $\varepsilon_0$ are 0. The initial state is fed into Crack-Net to predict the subsequent step's state, which is then used as input to predict the preceding steps' states. From Extended Data Table 1, it is evident that Crack-Net possesses the ability to predict the transformed increment, which is subsequently converted back to the absolute increment and added to the phase field in the current step to obtain the phase field for the next step. Additionally, the stress in the next step is directly predicted by Crack-Net. Given the fixed interval of strain ratio, the strain ratio in each step can be calculated, enabling acquisition of the state for the next step from Crack-Net's predictions. This process forms a loop that can be iterated until reaching the maximum value set beforehand. As observed from the long-term prediction process, Crack-Net can predict the crack phase field and stress at each step of the fracture process based on the given composite design. Remarkably, Crack-Net has the flexibility to predict from any step of the fracture by simply altering the initial state in Extended Data Table 1 to match the desired step's state.

**Transfer learning for Crack-Net**

Transfer learning is a machine learning technique in which knowledge gained from solving one task (the source domain) is utilized to improve the performance of another related task (the target domain)[43]. In transfer learning, a pre-trained model, which has been trained on a large dataset in the source domain, is adapted and fine-tuned for a target domain with a smaller dataset. The basic idea behind transfer learning is that the knowledge learned by a model in the source domain can be generalized and applied to the target domain, even if the target domain has a different dataset or distribution. By leveraging the knowledge from the source task, the model can start with a good initial configuration and then fine-tune its parameters on the target task to achieve better performance with less data and training time. In the context of Crack-Net, transfer learning plays a pivotal role in enhancing the model's predictive capabilities for composite materials with varying material properties. As illustrated in Fig. 4a, parameter transfer learning is adopted in this work, which utilizes the learned parameters of a pre-trained model from the source domain to initialize or fine-tune the model for the target domain. When applying transfer learning, the parameters of the pre-trained model are transferred to the target domain, and then the model is further trained using the target domain's data to adapt it to the specific characteristics of the new task. In this work, the entire pre-trained model, including both the encoder and decoder, is fine-tuned using the data in target domain. The parameters are updated based on the loss function in the target domain, allowing the model to adjust its representations and learn task-specific patterns. For the fine-tuning process, the training epoch is 1000, the learning rate is reduced to $10^{-4}$, the batch size is 100, and the early-stopping technique is employed to prevent overfitting. For the scenario with the strength of the reinforcement phase being reduced to UTS = 4.8 and $E_0$ = 100, then 50 cases (including 12,500 steps) with different composite designs are adopted as the dataset to fine-tune the model. During the training process, we partitioned the dataset into three sets, allocating 80% for training (10,000 steps), 10% for validation (1,250 steps), and another 10% for testing (1,250 steps). For the scenario with the strength of the reinforcement phase being increased to UTS = 112 and $E_0$ = 50,000, then 50 cases (including 5,000 steps) with different composite designs are adopted as the dataset to fine-tune the model. During the training process, we partitioned the dataset into three sets, allocating 80% for training (4,000 steps), 10% for validation (500 steps), and another 10% for testing (500 steps). The training epoch is 1,000 and the learning rate is reduced to $10^{-4}$.

During the training of transfer learning, a challenge arose from the significant differences in UTS and $E_0$ of different materials in the composites. Directly inputting these diverse material properties into the network during training could cause the batch normalization layer in Crack-Net to experience substantial changes in the mean and variance of the batch, which could adversely affect the effectiveness of transfer learning. To address this issue, a crucial preprocessing step was implemented for the material properties of the composite materials. Given the correlation between the mode of crack propagation and the relative strength of the matrix and reinforcement phase of the composite material, we applied normalization techniques to bring the material properties in the target domain closer to the order of magnitude in the source domain. By normalizing the material properties, we ensured that their values were scaled appropriately, enabling the transfer learning process to proceed more smoothly. The normalized material properties can be written as:

$$\text{UTS}_{\text{target}} = \text{UTS}_{\text{target}} \times \frac{\max(\text{UTS}_{\text{source}})}{\max(\text{UTS}_{\text{target}})}, \tag{7}$$

$$E_{0\,\text{target}} = E_{0\,\text{target}} \times \frac{\max(E_{0\,\text{source}})}{\max(E_{0\,\text{target}})}. \tag{8}$$

**Data availability**

The dataset generated in this study has been deposited in the GitHub repository, https://github.com/woshixuhao/Crack-Net/tree/main.

**Code availability**

All original code has been deposited in the GitHub repository, https://github.com/woshixuhao/Crack-Net/tree/main.

**Acknowledgements**

This work was supported and partially funded by the National Natural Science Foundation of China (Grant No. 52288101), the National Center for Applied Mathematics Shenzhen (NCAMS), the Shenzhen Key Laboratory of Natural Gas Hydrates (Grant No. ZDSYS20200421111201738), and the SUSTech – Qingdao New Energy Technology Research Institute.

**Author contributions**

W.F. constructed the numerical simulation to generate the fracture dataset. H.X. analyzed the data, constructed the deep learning model, and conducted studies to examine the model. H.X., W.F., L.R., and R.S conceived the idea. H.X. and W.F. designed the overall research. H.X. and W.F. wrote the manuscript. A.C.T. and D.Z. revised the manuscript and supervised the whole project. L.R. and R.S assisted handling the engineering problems in the project.

**Competing interests**

All authors declare no competing interests.

**Extended Data Table 1.** The algorithm for long-term prediction by Crack-Net

---

**Algorithm 1: Long-term prediction via Crack-Net**

---

Initial:      $(d_0, \sigma_0, \varepsilon_0)$, $\Delta\varepsilon=1\%$, UTS and $E_0$ are constant fields.

$1^{th}$ step:    $(d_0, \sigma_0, \varepsilon_0) \xrightarrow{\text{Crack-Net}} (\Delta\hat{d}'_0, \hat{\sigma}_1)$

$$\hat{d}_1 = d_0 + 10^{\Delta\hat{d}'_0 - 10}$$

$$\varepsilon_1 = \varepsilon_0 + \Delta\varepsilon$$

Get the prediction $(\hat{d}_1, \hat{\sigma}_1, \varepsilon_1)$

Loop:      $(\hat{d}_k, \hat{\sigma}_k, \varepsilon_k) \xrightarrow{\text{Crack-Net}} (\hat{d}_{k+1}, \hat{\sigma}_{k+1})$

$$\hat{d}_{k+1} = d_k + 10^{\Delta\hat{d}'_k - 10}$$

$$\varepsilon_{k+1} = \varepsilon_k + \Delta\varepsilon$$

Get the prediction $(\hat{d}_{k+1}, \hat{\sigma}_{k+1}, \varepsilon_{k+1})$

Until iteration $k$ achieves the maximum iteration $N$.

---

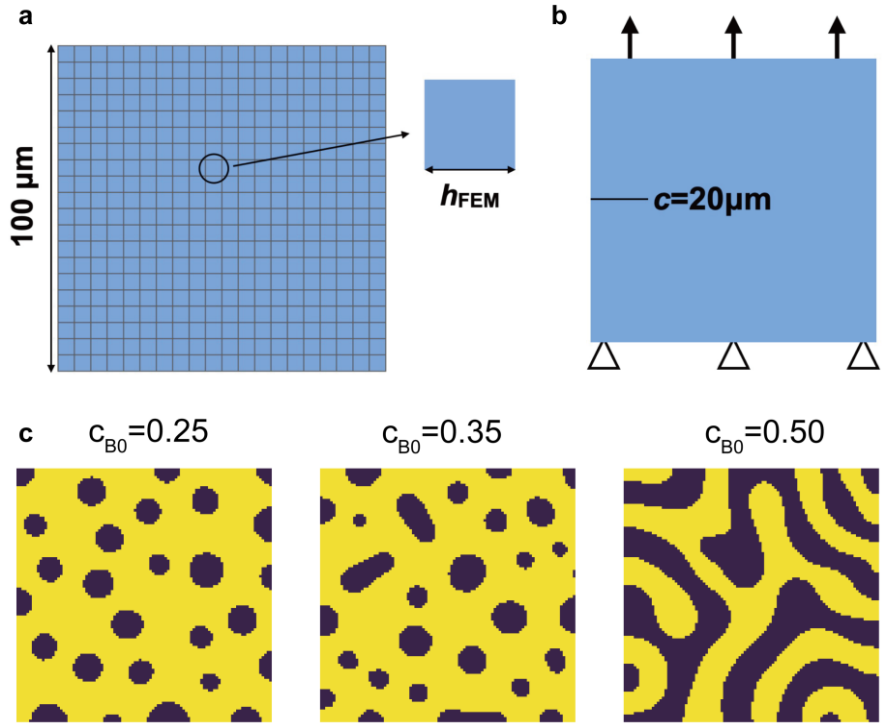

**Extended Data Figure 1. The schematic of simulation set-up. (a)** The meshing elements in the numerical simulation, $h_{FEM}$=1μm during simulations and $h_{FEM}$=5μm here for clear viewing. **(b)** The boundary conditions of simulations. **(c)** Examples of the generated morphology with $c_{B0}$ ranging from 0.25 to 0.5.

# Supplementary Information for

# Crack-Net: Prediction of Crack Propagation in Composites

Hao Xu[1,†], Wei Fan[2,†,*], Ambrose C. Taylor[2], Dongxiao Zhang[3,4,*], Lecheng Ruan[5], Rundong Shi[6]

[1] BIC-ESAT, ERE, and SKLTCS, College of Engineering, Peking University, Beijing 100871, P. R. China

[2] Department of Mechanical Engineering, Imperial College London, South Kensington Campus, London, SW7 2AZ, UK

[3] Eastern Institute for Advanced Study, Eastern Institute of Technology, Ningbo, Zhejiang 315200, P. R. China

[4] Department of Mathematics and Theories, Peng Cheng Laboratory, Shenzhen 518000, Guangdong, P. R. China

[5] Department of Mechanical and Aerospace Engineering, University of California, Los Angeles, Los Angeles, CA, 90095, USA

[6] Meituan Corporate

* Corresponding authors. Email address: w.fan19@imerpial.ac.uk (W. Fan); dzhang@eias.ac.cn (D. Zhang).

† These authors contributed equally.

# 1. Supplementary Text
## 1.1 Out-of-sample long-term prediction of Crack-Net

In this work, Crack-Net is trained by the dataset of the composites with different structure designs. In order to truly reflect the generalization ability of Crack-Net, we conducted out-of-sample prediction in which several composites with new structure designs are generated for long-term prediction. Notably, these composite designs have never appeared in the training dataset, which means that they are data outside of the sample, and thus, the testing of these data is closer to the practical scenario. In Fig. 3 of the main text, we provided four examples with classic characteristics to demonstrate the performance of Crack-Net for long-term prediction faced with different types of composite designs. In order to improve the reliability of the conclusion, more out-of-sample prediction cases are provided in this section. In Fig. S1 to Fig. S15, the out-of-sample long-prediction by Crack-Net and the reference from numerical simulation under different composite designs are depicted. Our Crack-Net is capable of concurrently producing both the stress-strain curve and the corresponding crack phase field at each step. Given that the complete crack propagation process in composite materials often encompasses numerous steps, we have chosen to illustrate a subset of these stages within the figure due to spatial constraints of the figure. The selected stages span the initial, intermediate, and final phases of crack propagation, offering a representative portrayal. For a comprehensive visualization of the predicted full fracture process, some .gif diagrams are provided in open source. From the figures, it is observed that the prediction of Crack-Net matches well with the reference from numerical simulation in most of the cases, which confirms that Crack-Net achieves a satisfactory performance in the long-term prediction of crack phase field and stress-strain curve for composites with different composite designs. Notably, in certain instances, subtle distinctions between the predicted crack phase field by Crack-Net and the reference in the same step occurred. This is because Crack-Net occasionally forecasted an earlier initiation of fracture. As discussed in the main text, Crack-Net learned the relationship between stress development and crack growth during the crack propagation process. Therefore, in the long-term prediction of Crack-Net, the evolution of the crack phase field aligns with the stress-strain curve. In practical applications, more focus is usually given to the fracture characteristics of composite materials instead of specific steps. Therefore, although Crack-Net sometimes predicts the emergence of cracks earlier, this predictive discrepancy remains inconsequential since Crack-Net captures the characteristics of the crack propagation. This is particularly evident in Fig. 3 of the main text, where Crack-Net's forecasted stress-strain curve characteristics harmonize with observed patterns, and the anticipated crack phase field at the characteristic steps is parallel with the reference from numerical simulation. Meanwhile, it is discovered that the development process and final morphology of cracks are predicted accurately in these cases, which confirms that Crack-Net has learned the characteristics of the crack propagation mechanism.

Among the out-of-sample prediction cases, the crack phase field prediction in Fig. S13 to S15 shows minor deviations compared with the reference. Here, we will analyze the causes and impacts of these deviations in detail. The comparison between the predicted and observed ultimate crack phase field for these cases is displayed in Fig. S16 for better demonstration. In Fig. S16a, an uncommon phenomenon in the observed crack phase field occurs in which cracks are discontinuous, with a separate segment of the crack appearing at the boundary, which poses a great challenge for crack propagation prediction. While the discontinuous cracks at boundaries were not explicitly identified, Crack-Net demonstrated remarkable precision in predicting the advancement trajectory of the primary crack. In Fig. S16b, the terminal end of the predicted crack phase field displays a noticeable deviation from the observation. However, the crack propagation process predicted by Crack-Net adheres to physical principles. Given the lower initial elastic modulus ($E_0$) and ultimate tensile strength (UTS) of the reinforcing phase compared to the matrix, cracks tend to propagate through the reinforcement phase. Notably, at the terminal end of the crack, two potential options emerge – traversing the particle reinforcement phase material either from the upper right (reference) or the lower right (prediction). Both scenarios align with physical laws, and in practical experiments, these two instances could plausibly transpire due to minor disturbances. In this context, Crack-Net presents a feasible crack propagation process that could manifest in real-world scenarios. In Fig. S16c, the disparity between the prediction and observation is subtle, and is primarily concentrated in the initial part of the crack path. Likewise, the crack propagation process predicted by Crack-Net is consistent with the previously discussed physical laws, i.e., cracks tend to expand in more brittle materials. However, Crack-Net provides another reasonable possibility for the path through the material. In summary, a large number of examples have proven that Crack-Net achieves good accuracy in long-term prediction of crack propagation patterns, which satisfies physical laws, even though a minority of examples exhibit disparities from actual crack phase fields.

In the main text, the generalization ability of Crack-Net to a more complex composite design, i.e., the two-phase co-continuous structure, is examined, and Fig. 3 of the main text provides an illustrative example. In this section, we provide more examples to make the claim more convincing. Fig. S17 to Fig. S19 depicts the long-term prediction by Crack-Net for two-phase co-continuous composite materials with different composite designs. It is found that Crack-Net exhibits remarkable efficacy in predicting crack propagation, even in the complex scenario of two-phase co-continuous structures. Such structures, characterized by distinct reinforcement phases interwoven in a continuous matrix, present intricate crack propagation pathways that can be challenging to predict accurately. However, Crack-Net can handle this challenge, as exemplified by its predictions aligning well with actual crack phase fields and stress-strain curves of such structures.

**1.2 Additional experiments for the transfer learning of Crack-Net**
In this work, the transfer learning technique is adopted to enable Crack-Net to predict the crack propagation process of the composites with both different composite designs and material attributes. Fig. 4 in the main text has proven that transfer learning is effective in predicting the crack phase field and stress-strain curve of the composites with different reinforcement strengths, even though the mechanism of crack development is completely different. In the main text, 120 cases are utilized as the dataset, and the result of transfer learning is apparently more accurate than the model without transfer learning, demonstrating satisfactory predictive ability. Considering that the Crack-Net in the source domain is trained with 554 cases, the size of the dataset utilized in the transfer learning is much smaller than that in the source domain, which confirms that this approach is flexible and applicable. In this section, we conduct an additional experiment to examine the performance of Crack-Net with different amounts of data for transfer learning. In the experiment, the dataset with 25, 50, and 120 cases are employed for transfer learning, respectively, and the results are shown in Fig. S20. Here, the strength of the reinforcement phase is UTS=112 and $E_0$=50,000. From the figure, it is observed that the augmentation of the dataset for transfer learning directly correlates with an escalation in the predictive accuracy of the model, since the model trained with 120 cases in transfer learning exhibits a perfect match with the reference. Meanwhile, it is revealed that the stress-strain curve is easier to be transferred with a small number of cases in transfer learning, in which the model transferred with 25 cases can predict the pattern of stress-strain curve well, but fail to provide a correct crack phase field. This confirms that the long-term prediction of crack phase field propagation is a more challenging task than predicting a one-dimensional stress-strain curve, which reflects the significance of Crack-Net. Meanwhile, it is revealed that when the data size is reduced to 50, the Crack-Net still has good predictive performance.

In order to better demonstrate the superiority of the transfer learning, we conduct another additional experiment by comparing the performance of Crack-Net directly trained by the same dataset utilized in the transfer learning. Here, the data volume is selected to be 50. In direct training, we utilize this dataset to initiate model training from the ground up. We focus on two typical cases in which the strength of the reinforcement phase is large and small, respectively, as demonstrated in the main text. The comparison is depicted in Fig. S21. From the figure, it is discovered that the predictive ability of the model directly trained by the dataset without transfer learning is generally unsatisfactory in both situations, since it failed to the predict the correct crack phase field. In contrast, the Crack-Net trained by transfer learning exhibited commendable accuracy. This confirms that the learned underlying mechanism of Crack-Net in the source domain can facilitate the training of the model in the target domain, which makes the transfer learning effective in this task.

In the main text, it is claimed that the Crack-Net model possesses good predictive ability for composites with reinforcement strength within a certain range. In order to further confirm the explanation, we employ Crack-Net to predict crack propagation with a slightly different reinforcement strength. In the experiment, we employ the Crack-Net trained with data from composites with a weak reinforcement phase (UTS=10.8, $E_0$=500) to directly predict the crack propagation in the composites with a slightly weaker reinforcement phase (UTS=9.63, $E_0$=400). Meanwhile, we employ the Crack-Net trained with data from composites with a strong reinforcement phase (UTS=112, $E_0$=50,000) by transfer learning to directly predict the crack propagation in the composites with a slightly weaker reinforcement phase (UTS=100, $E_0$=40,000). The results are shown in Fig. S22, which depicts two cases, in which the composite design remains the same with the reinforcement strength is changed in each case. From the figure, it is evident that Crack-Net is able to provide satisfactory prediction for the composites with slightly changed reinforcement strength, since the predicted ultimate crack phase field is correct and the stress-strain curve pattern is captured. Notably, it is confirmed that the propagation of crack in composite materials is notably influenced by the relative strength of the matrix and reinforcement phase (Fig. S22) since the crack growth is divergent in the composite with the same composite design and different reinforcement strength.

## 2. Supplementary Figures and Tables

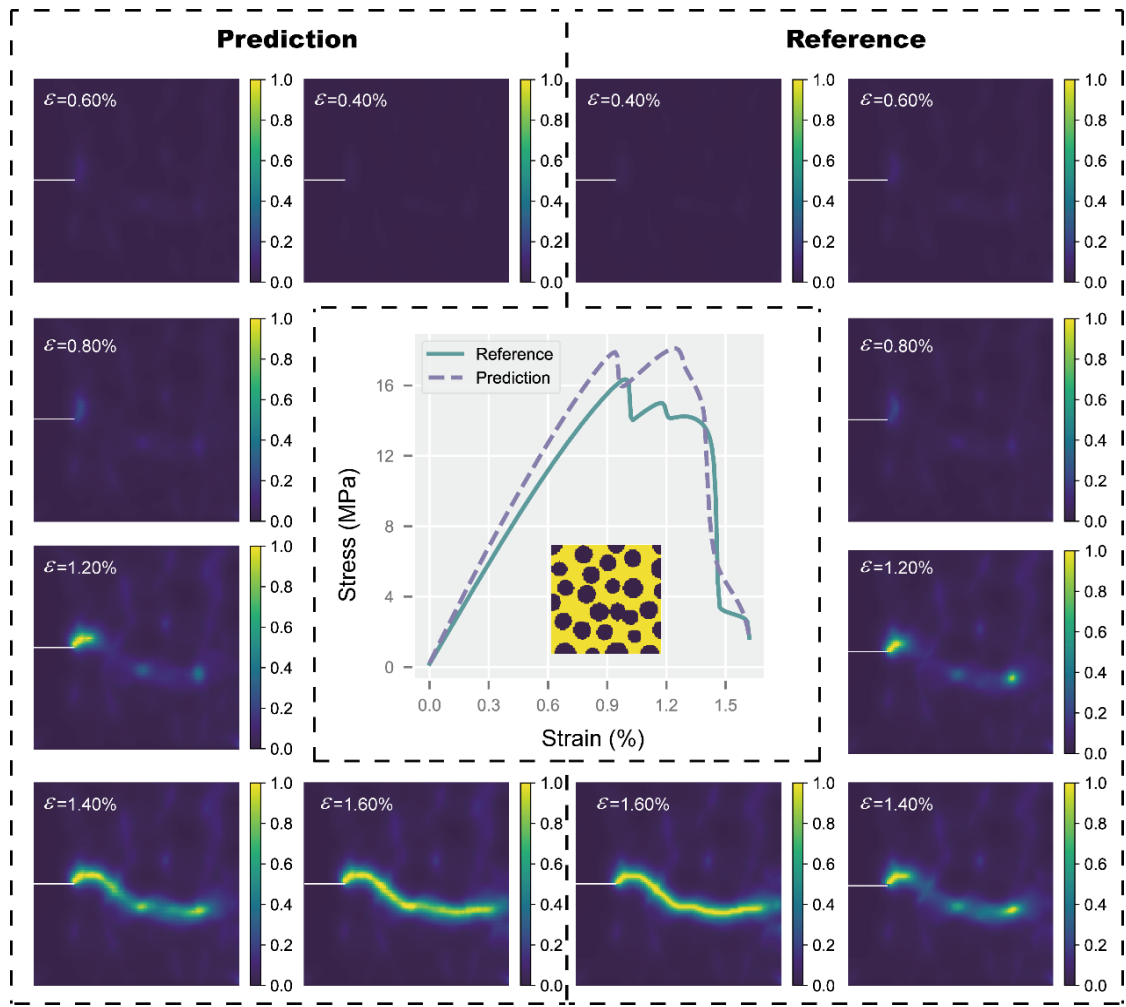

**Fig. S1. The out-of-sample long-prediction by Crack-Net and the reference from numerical simulation.** The center of the figure is the predicted stress-strain curve and the reference, and the composite design is displayed at the bottom. In the composite design, the yellow region refers to the matrix, and the black region refers to the reinforcement phase. On the left side of the figure, predicted phase fields for various steps are displayed, while the right side exhibits the actual phase fields in corresponding steps obtained from numerical simulation. The white line in the phase field represents the initial condition.

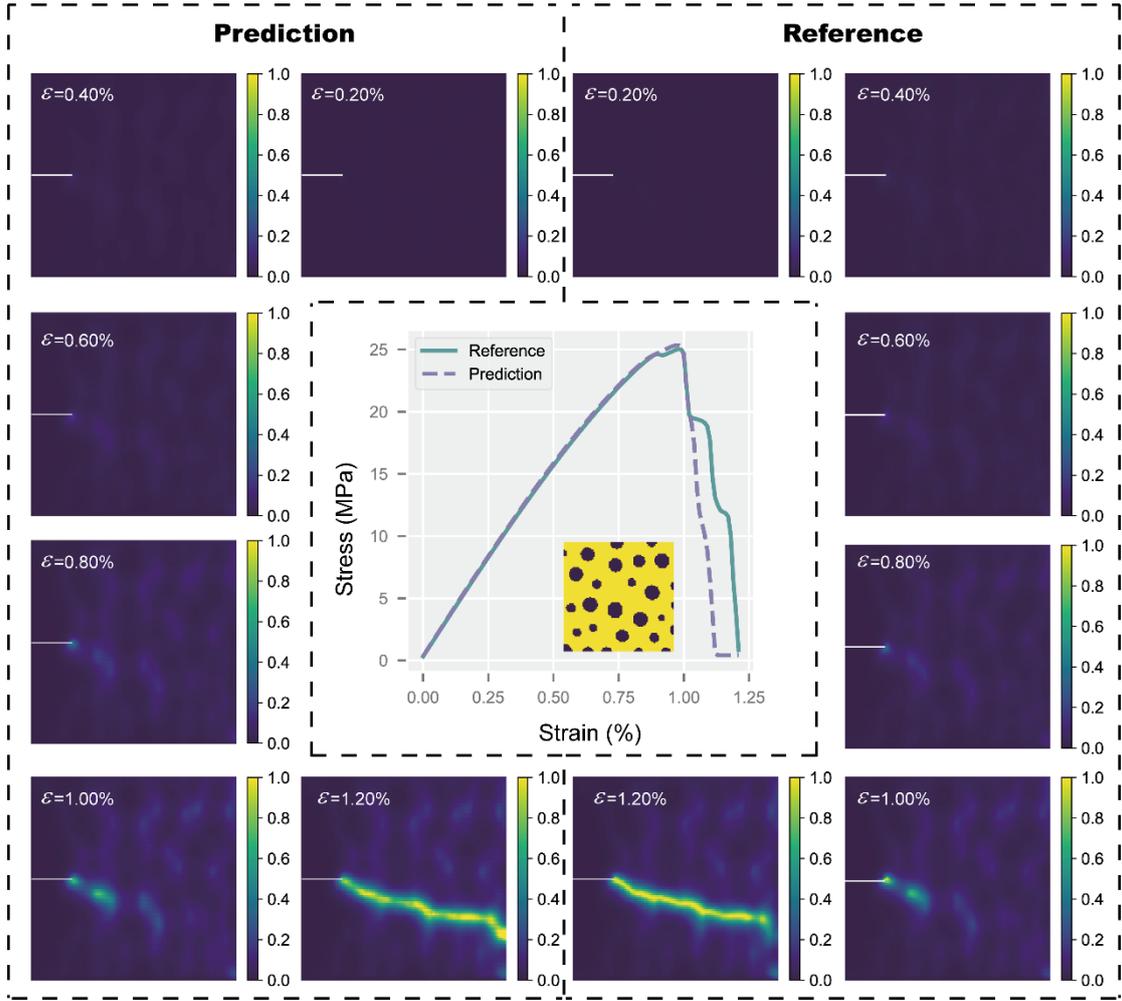

**Fig. S2. The out-of-sample long-prediction by Crack-Net and the reference from numerical simulation.** The center of the figure is the predicted stress-strain curve and the reference, and the composite design is displayed at the bottom. In the composite design, the yellow region refers to the matrix, and the black region refers to the reinforcement phase. On the left side of the figure, predicted phase fields for various steps are displayed, while the right side exhibits the actual phase fields in corresponding steps obtained from numerical simulation. The white line in the phase field represents the initial condition.

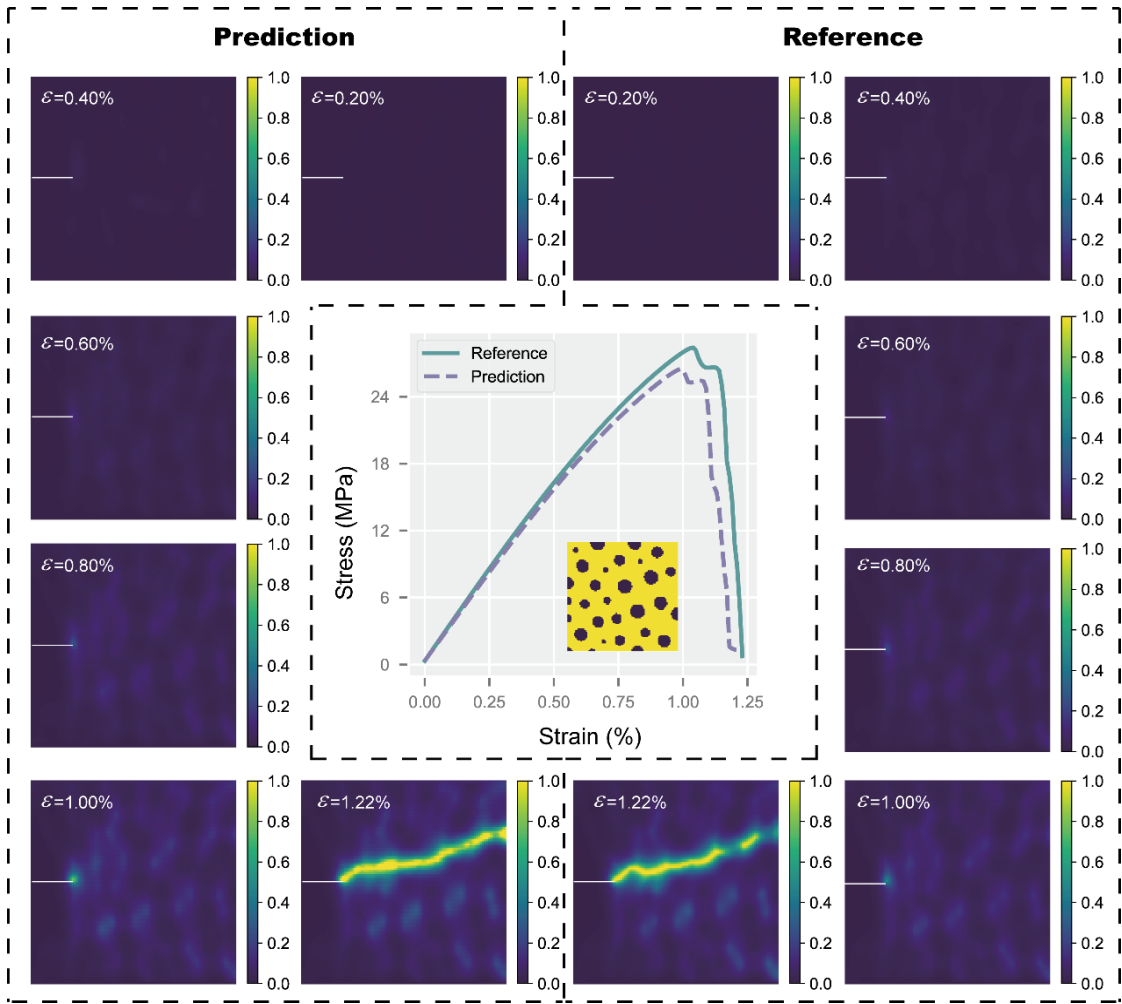

**Fig. S3. The out-of-sample long-prediction by Crack-Net and the reference from numerical simulation.** The center of the figure is the predicted stress-strain curve and the reference, and the composite design is displayed at the bottom. In the composite design, the yellow region refers to the matrix, and the black region refers to the reinforcement phase. On the left side of the figure, predicted phase fields for various steps are displayed, while the right side exhibits the actual phase fields in corresponding steps obtained from numerical simulation. The white line in the phase field represents the initial condition.

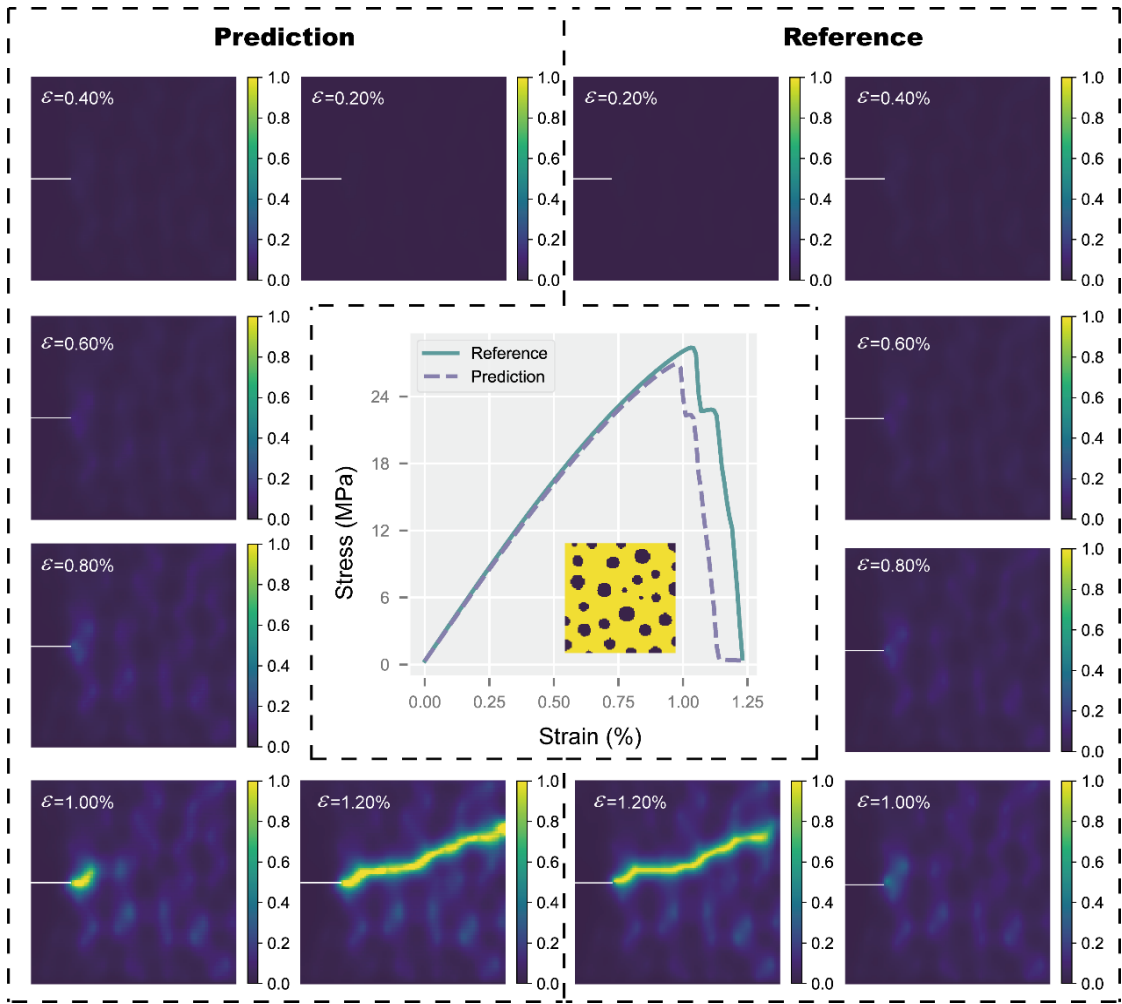

**Fig. S4. The out-of-sample long-prediction by Crack-Net and the reference from numerical simulation.** The center of the figure is the predicted stress-strain curve and the reference, and the composite design is displayed at the bottom. In the composite design, the yellow region refers to the matrix, and the black region refers to the reinforcement phase. On the left side of the figure, predicted phase fields for various steps are displayed, while the right side exhibits the actual phase fields in corresponding steps obtained from numerical simulation. The white line in the phase field represents the initial condition.

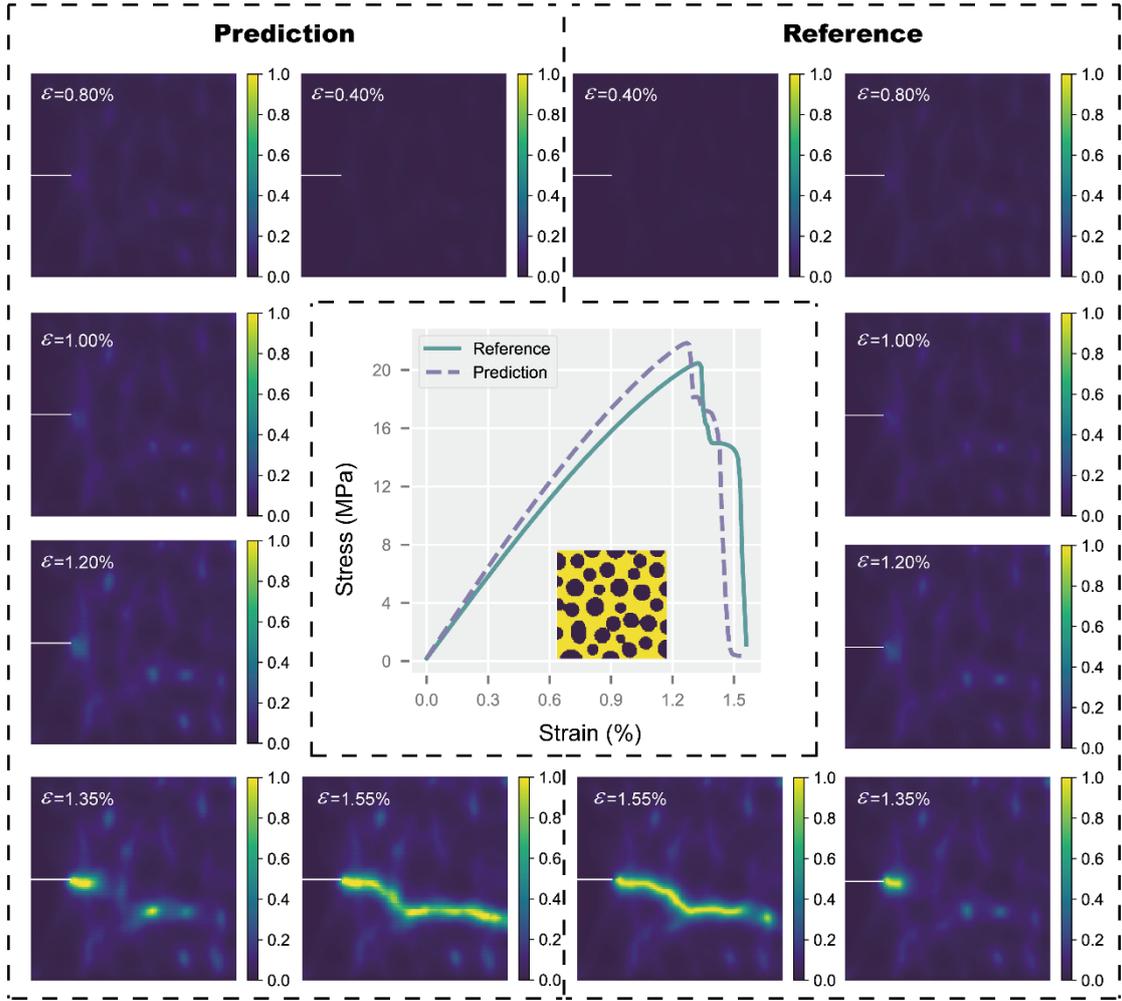

**Fig. S5. The out-of-sample long-prediction by Crack-Net and the reference from numerical simulation.** The center of the figure is the predicted stress-strain curve and the reference, and the composite design is displayed at the bottom. In the composite design, the yellow region refers to the matrix, and the black region refers to the reinforcement phase. On the left side of the figure, predicted phase fields for various steps are displayed, while the right side exhibits the actual phase fields in corresponding steps obtained from numerical simulation. The white line in the phase field represents the initial condition.

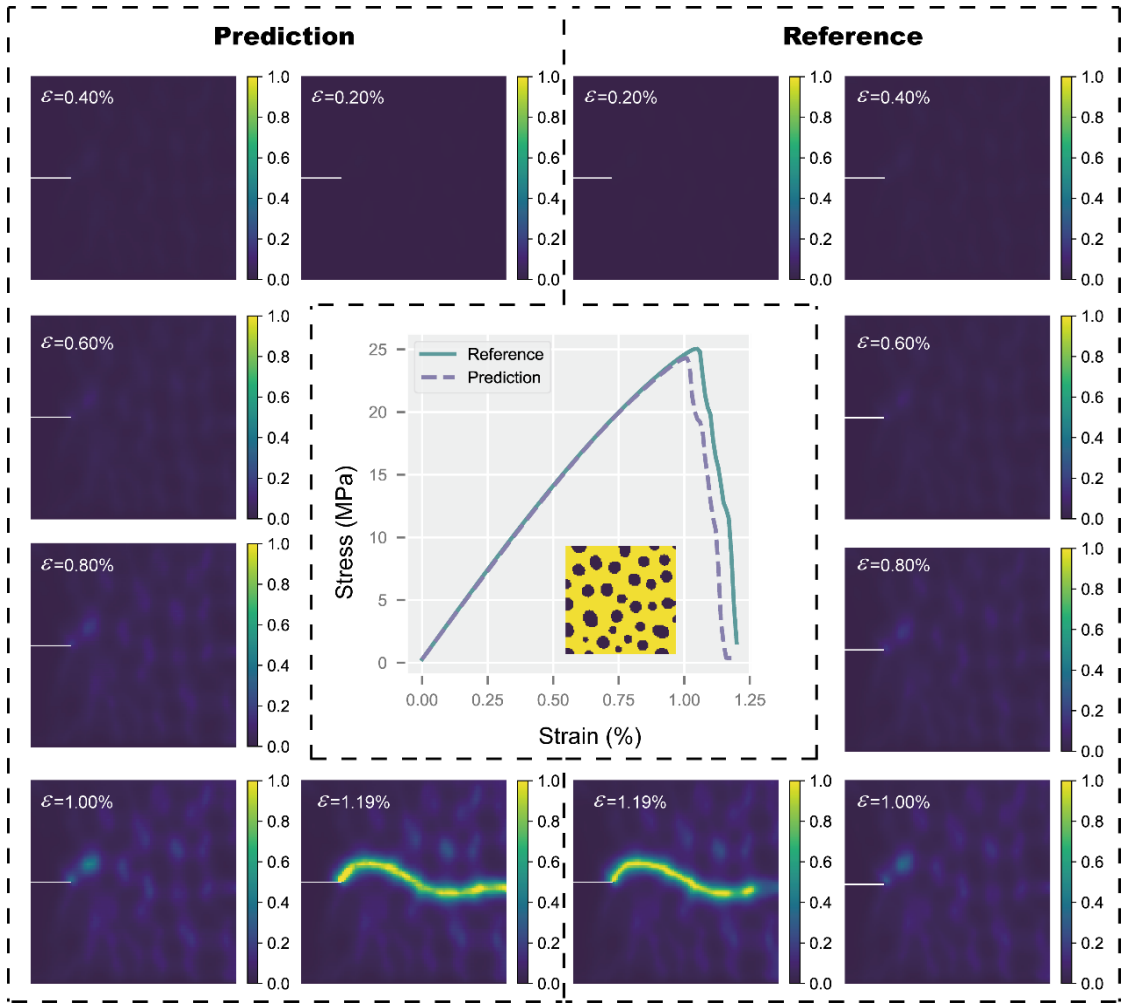

**Fig. S6. The out-of-sample long-prediction by Crack-Net and the reference from numerical simulation.** The center of the figure is the predicted stress-strain curve and the reference, and the composite design is displayed at the bottom. In the composite design, the yellow region refers to the matrix, and the black region refers to the reinforcement phase. On the left side of the figure, predicted phase fields for various steps are displayed, while the right side exhibits the actual phase fields in corresponding steps obtained from numerical simulation. The white line in the phase field represents the initial condition.

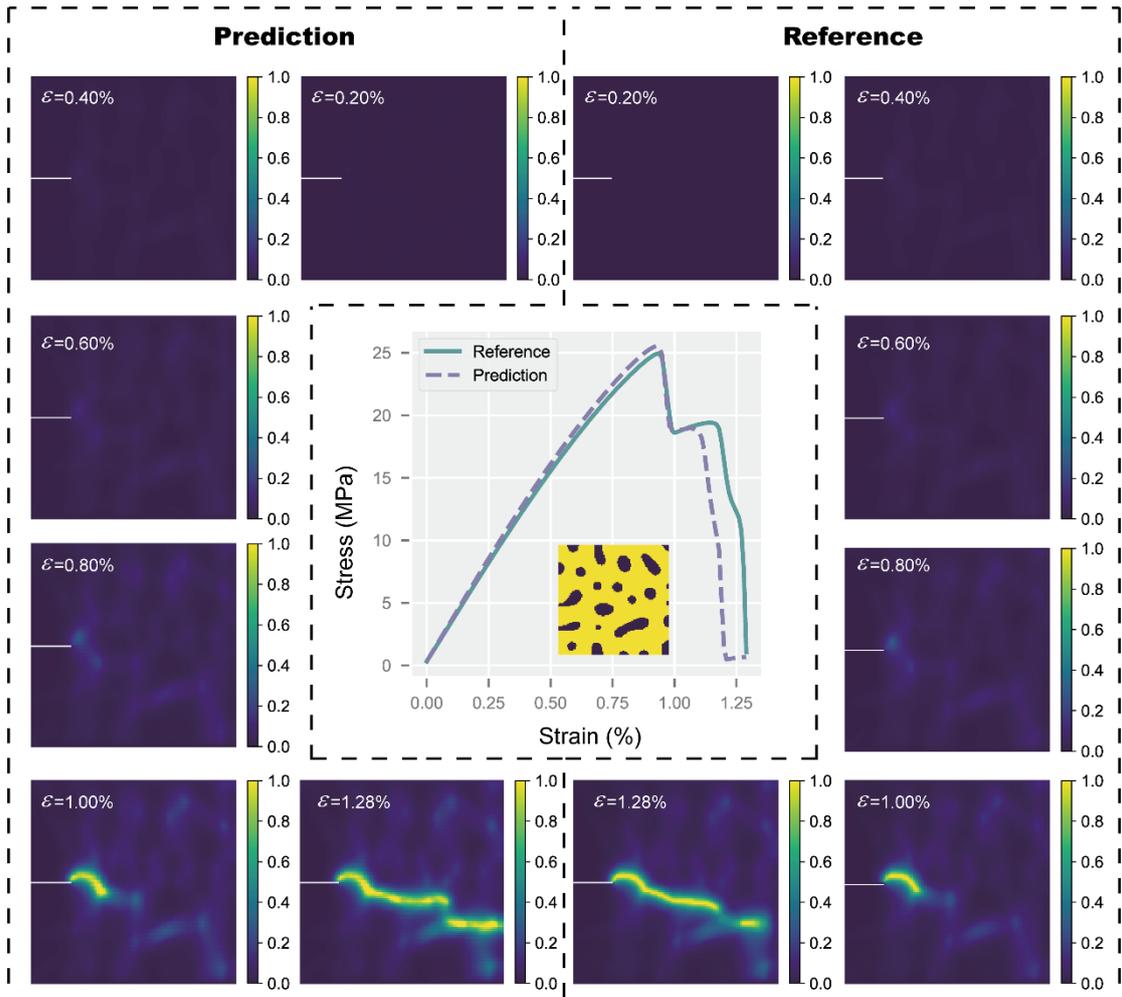

**Fig. S7. The out-of-sample long-prediction by Crack-Net and the reference from numerical simulation.** The center of the figure is the predicted stress-strain curve and the reference, and the composite design is displayed at the bottom. In the composite design, the yellow region refers to the matrix, and the black region refers to the reinforcement phase. On the left side of the figure, predicted phase fields for various steps are displayed, while the right side exhibits the actual phase fields in corresponding steps obtained from numerical simulation. The white line in the phase field represents the initial condition.

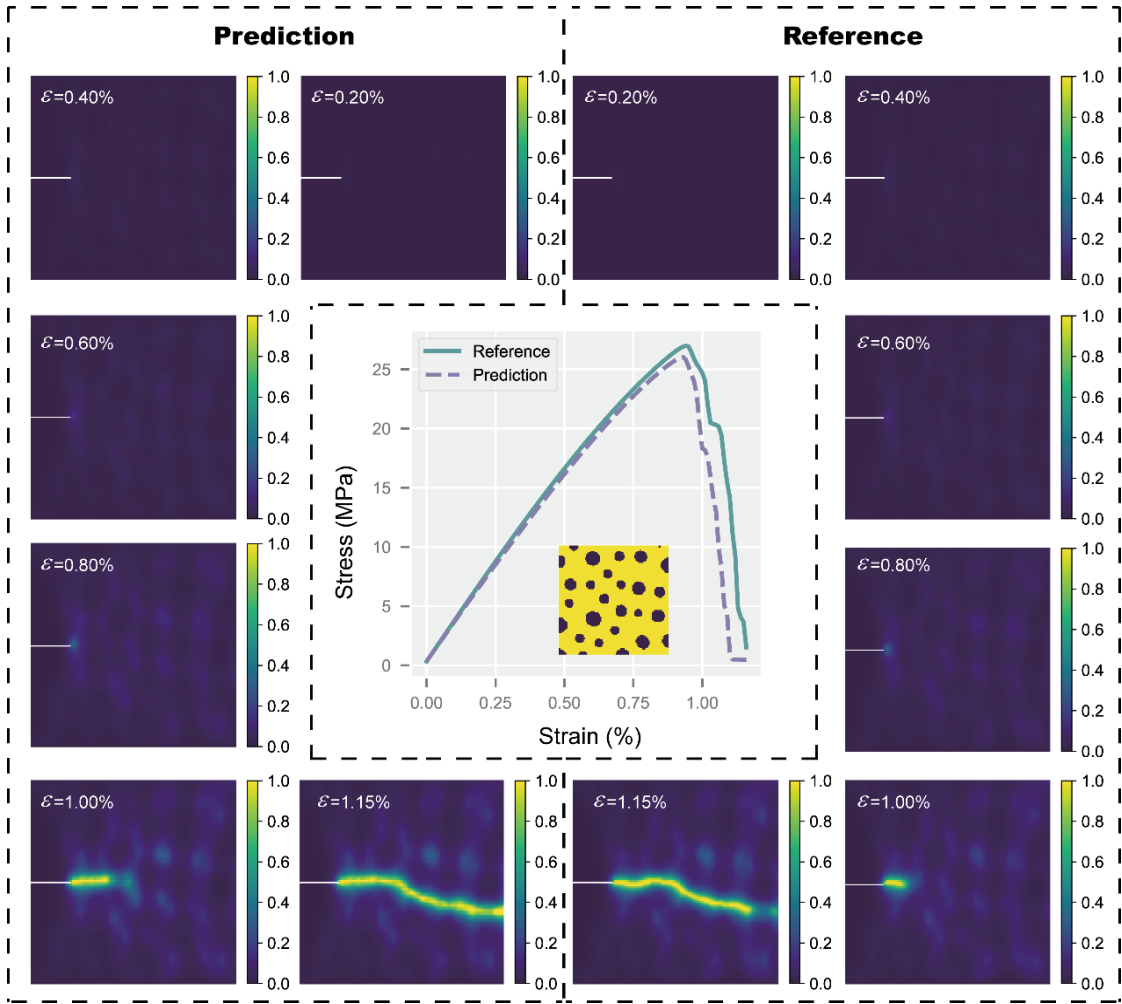

**Fig. S8. The out-of-sample long-prediction by Crack-Net and the reference from numerical simulation.** The center of the figure is the predicted stress-strain curve and the reference, and the composite design is displayed at the bottom. In the composite design, the yellow region refers to the matrix, and the black region refers to the reinforcement phase. On the left side of the figure, predicted phase fields for various steps are displayed, while the right side exhibits the actual phase fields in corresponding steps obtained from numerical simulation. The white line in the phase field represents the initial condition.

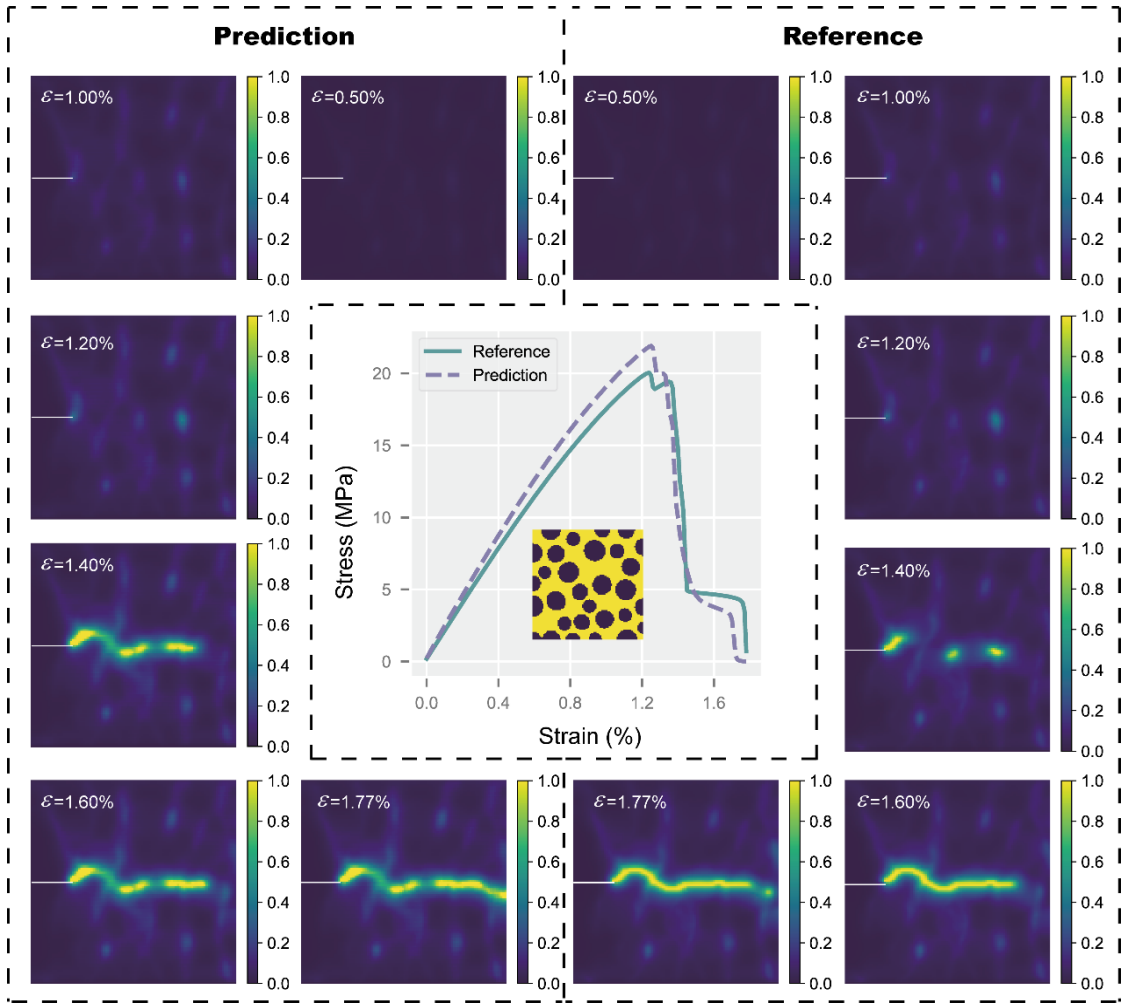

**Fig. S9. The out-of-sample long-prediction by Crack-Net and the reference from numerical simulation.** The center of the figure is the predicted stress-strain curve and the reference, and the composite design is displayed at the bottom. In the composite design, the yellow region refers to the matrix, and the black region refers to the reinforcement phase. On the left side of the figure, predicted phase fields for various steps are displayed, while the right side exhibits the actual phase fields in corresponding steps obtained from numerical simulation. The white line in the phase field represents the initial condition.

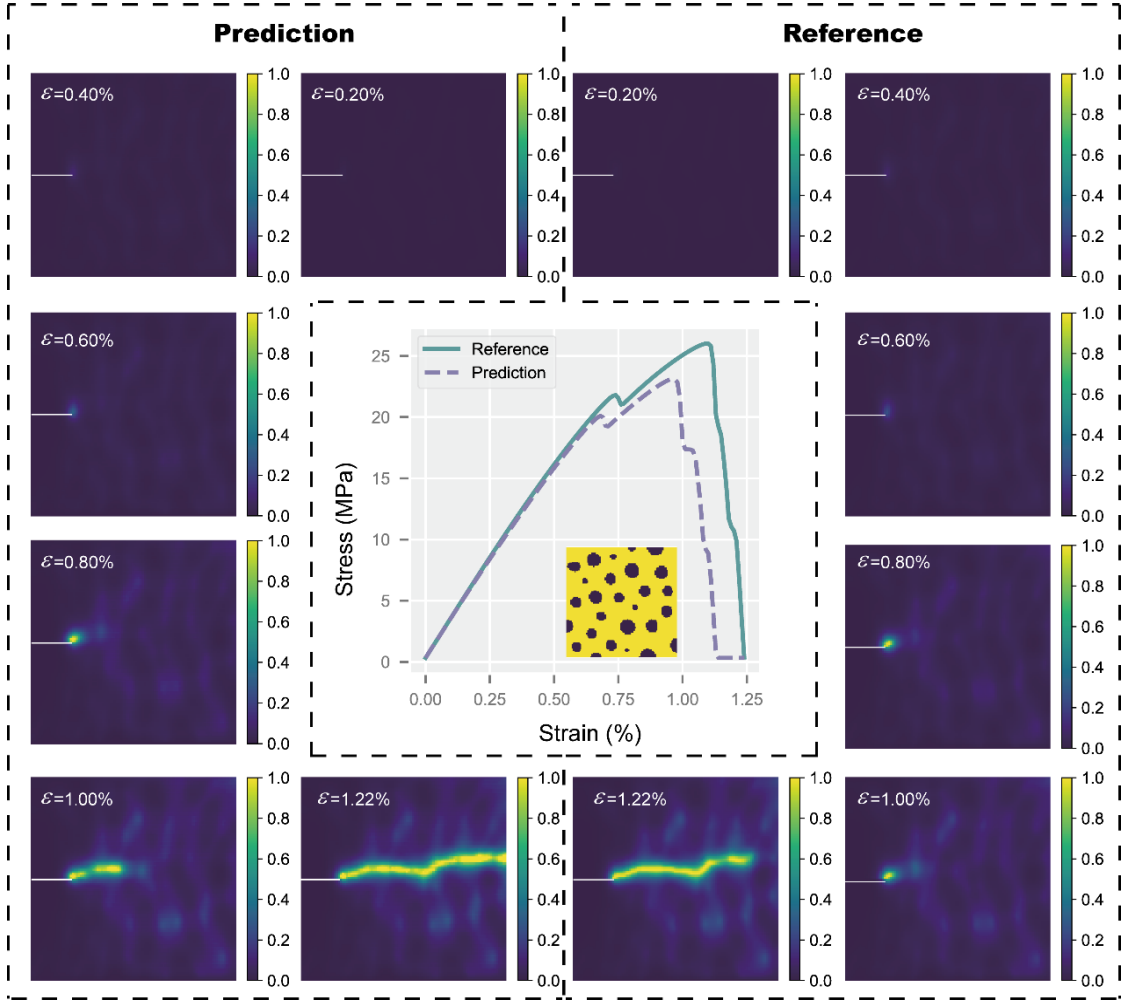

**Fig. S10. The out-of-sample long-prediction by Crack-Net and the reference from numerical simulation.** The center of the figure is the predicted stress-strain curve and the reference, and the composite design is displayed at the bottom. In the composite design, the yellow region refers to the matrix, and the black region refers to the reinforcement phase. On the left side of the figure, predicted phase fields for various steps are displayed, while the right side exhibits the actual phase fields in corresponding steps obtained from numerical simulation. The white line in the phase field represents the initial condition.

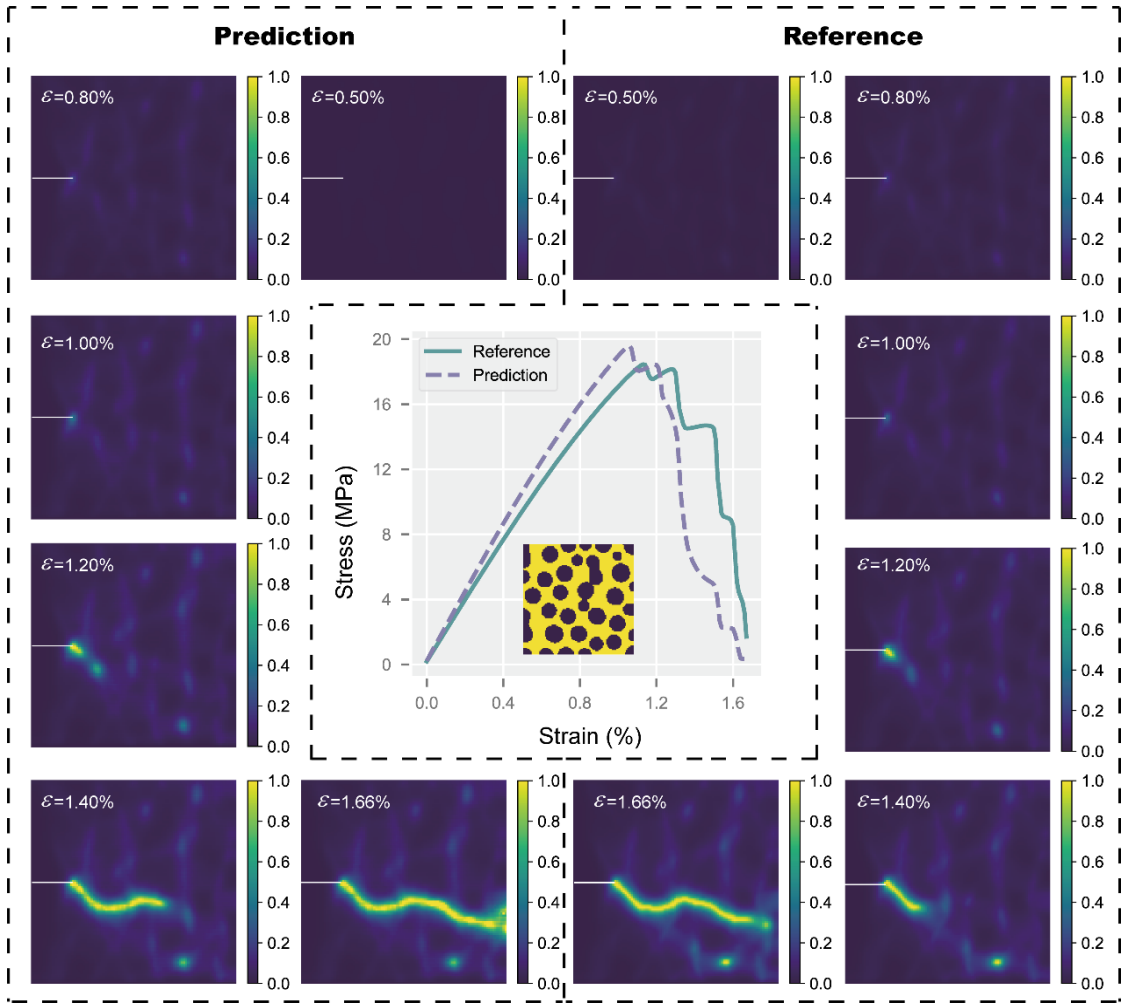

**Fig. S11. The out-of-sample long-prediction by Crack-Net and the reference from numerical simulation.** The center of the figure is the predicted stress-strain curve and the reference, and the composite design is displayed at the bottom. In the composite design, the yellow region refers to the matrix, and the black region refers to the reinforcement phase. On the left side of the figure, predicted phase fields for various steps are displayed, while the right side exhibits the actual phase fields in corresponding steps obtained from numerical simulation. The white line in the phase field represents the initial condition.

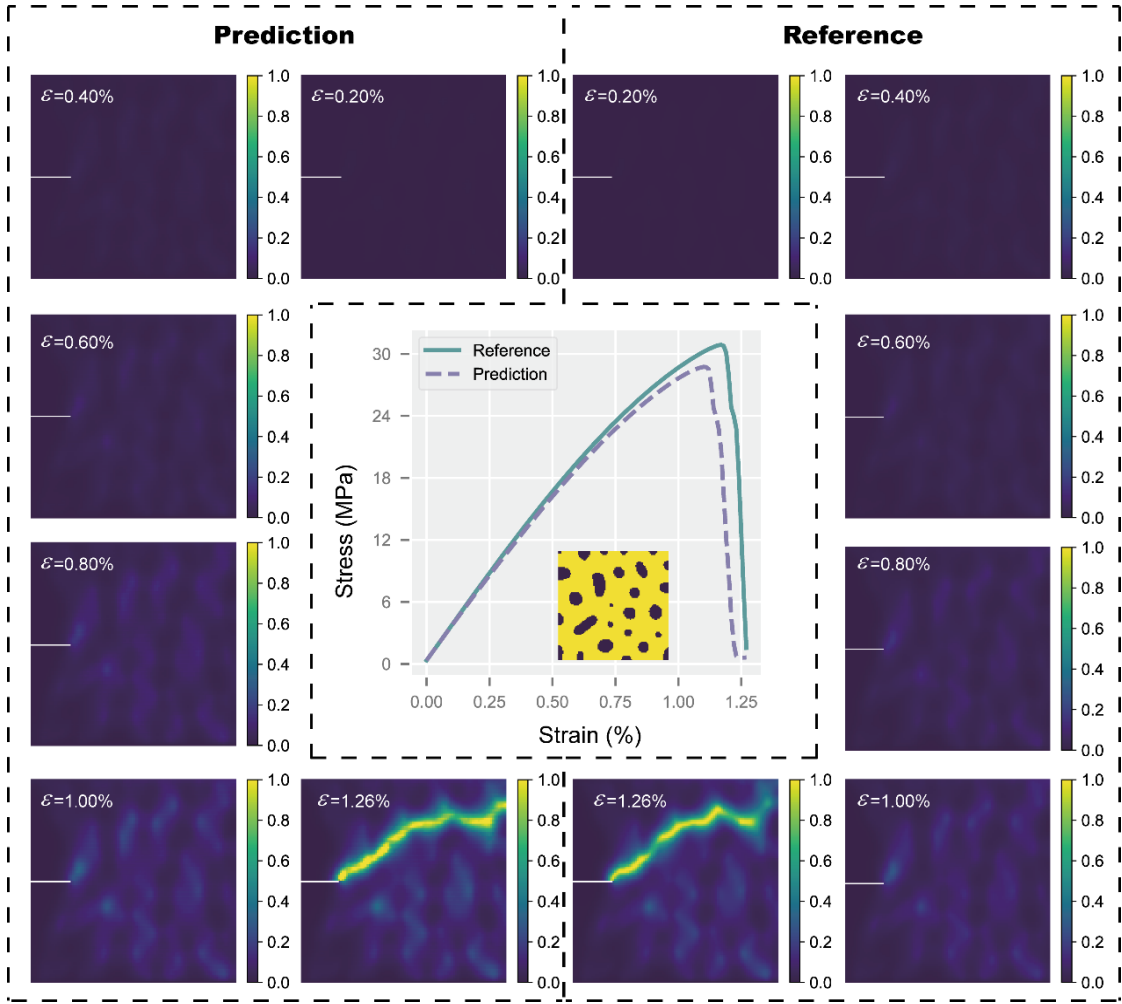

**Fig. S12. The out-of-sample long-prediction by Crack-Net and the reference from numerical simulation.** The center of the figure is the predicted stress-strain curve and the reference, and the composite design is displayed at the bottom. In the composite design, the yellow region refers to the matrix, and the black region refers to the reinforcement phase. On the left side of the figure, predicted phase fields for various steps are displayed, while the right side exhibits the actual phase fields in corresponding steps obtained from numerical simulation. The white line in the phase field represents the initial condition.

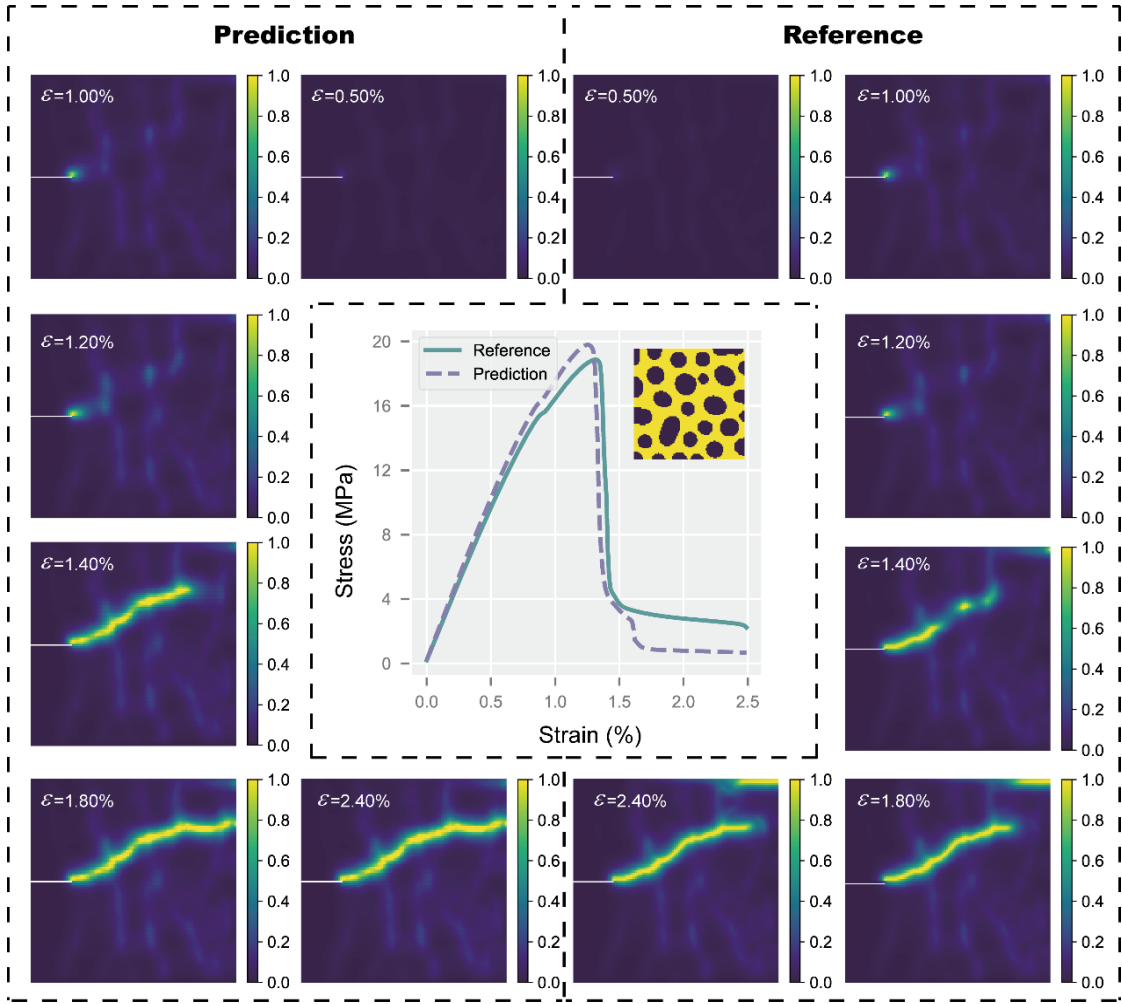

**Fig. S13. The out-of-sample long-prediction by Crack-Net and the reference from numerical simulation.** The center of the figure is the predicted stress-strain curve and the reference, and the composite design is displayed at the bottom. In the composite design, the yellow region refers to the matrix, and the black region refers to the reinforcement phase. On the left side of the figure, predicted phase fields for various steps are displayed, while the right side exhibits the actual phase fields in corresponding steps obtained from numerical simulation. The white line in the phase field represents the initial condition.

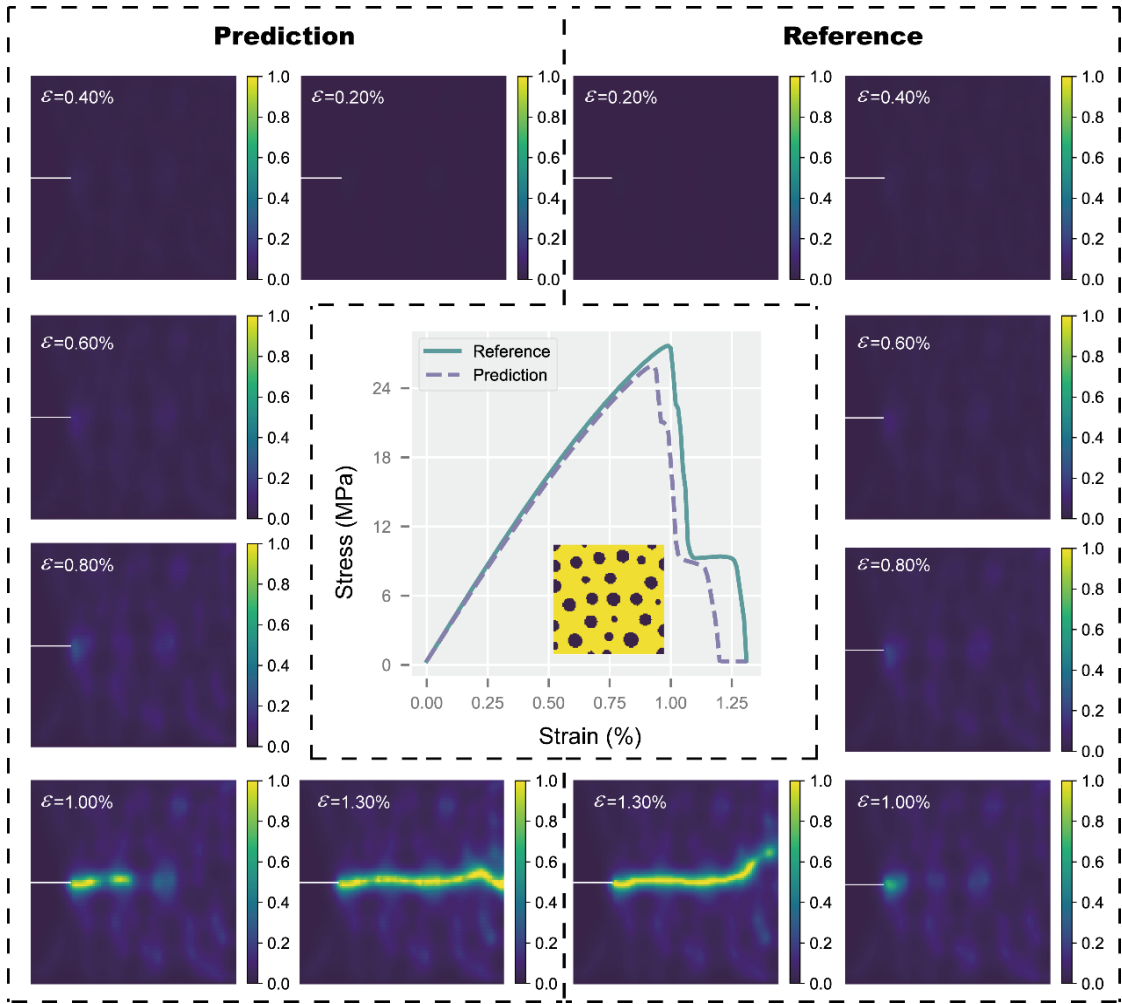

**Fig. S14. The out-of-sample long-prediction by Crack-Net and the reference from numerical simulation.** The center of the figure is the predicted stress-strain curve and the reference, and the composite design is displayed at the bottom. In the composite design, the yellow region refers to the matrix, and the black region refers to the reinforcement phase. On the left side of the figure, predicted phase fields for various steps are displayed, while the right side exhibits the actual phase fields in corresponding steps obtained from numerical simulation. The white line in the phase field represents the initial condition.

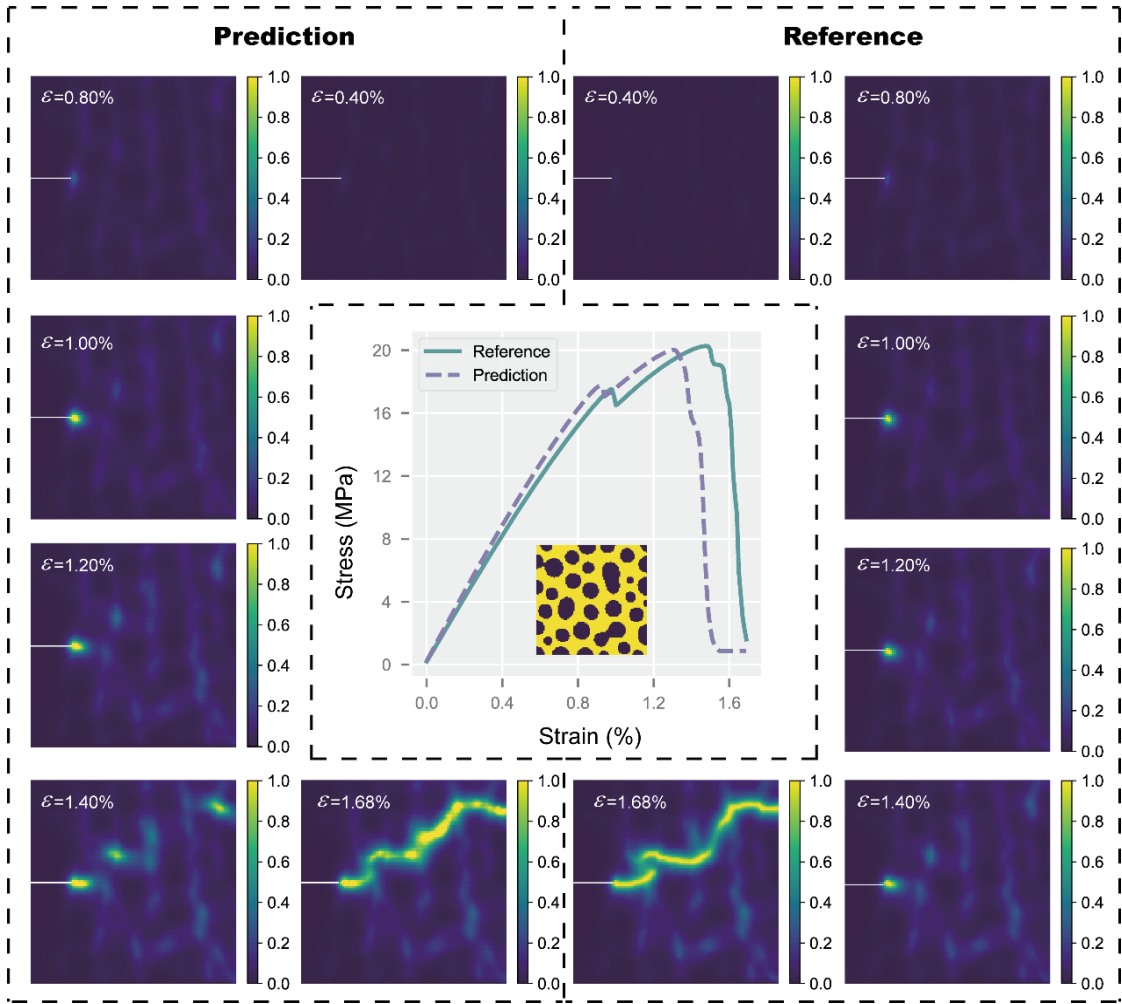

**Fig. S15. The out-of-sample long-prediction by Crack-Net and the reference from numerical simulation.** The center of the figure is the predicted stress-strain curve and the reference, and the composite design is displayed at the bottom. In the composite design, the yellow region refers to the matrix, and the black region refers to the reinforcement phase. On the left side of the figure, predicted phase fields for various steps are displayed, while the right side exhibits the actual phase fields in corresponding steps obtained from numerical simulation. The white line in the phase field represents the initial condition.

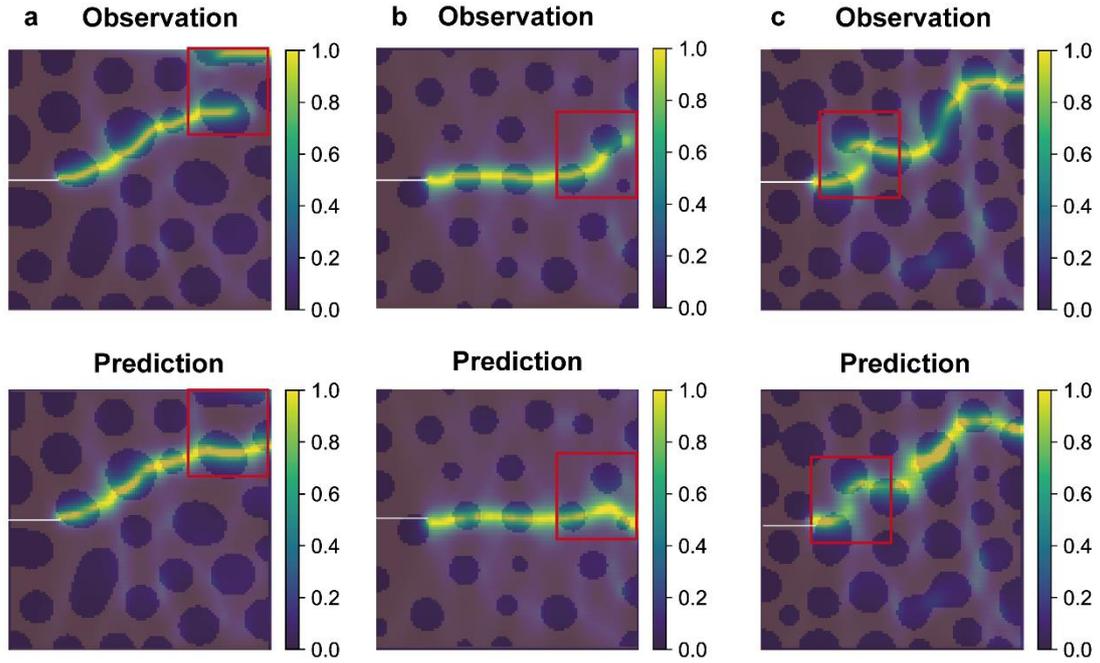

**Fig. S16. The comparison between the predicted and observed crack phase field for cases that exhibit minor deviation in long-term prediction.** **(a)** The prediction and observation of the ultimate crack phase field (strain=2.40%) for the case in Fig. S12. **(b)** The prediction and observation of the ultimate crack phase field (strain=1.30%) for the case in Fig. S13. **(c)** The prediction and observation of the ultimate crack phase field (strain=1.68%) for the case in Fig. S14. Deviations are highlighted within the red boxes. The crack phase field is overlaid with transparent composites design, in which the circular region represents the reinforcement phase. The white line in the phase field represents the initial condition.

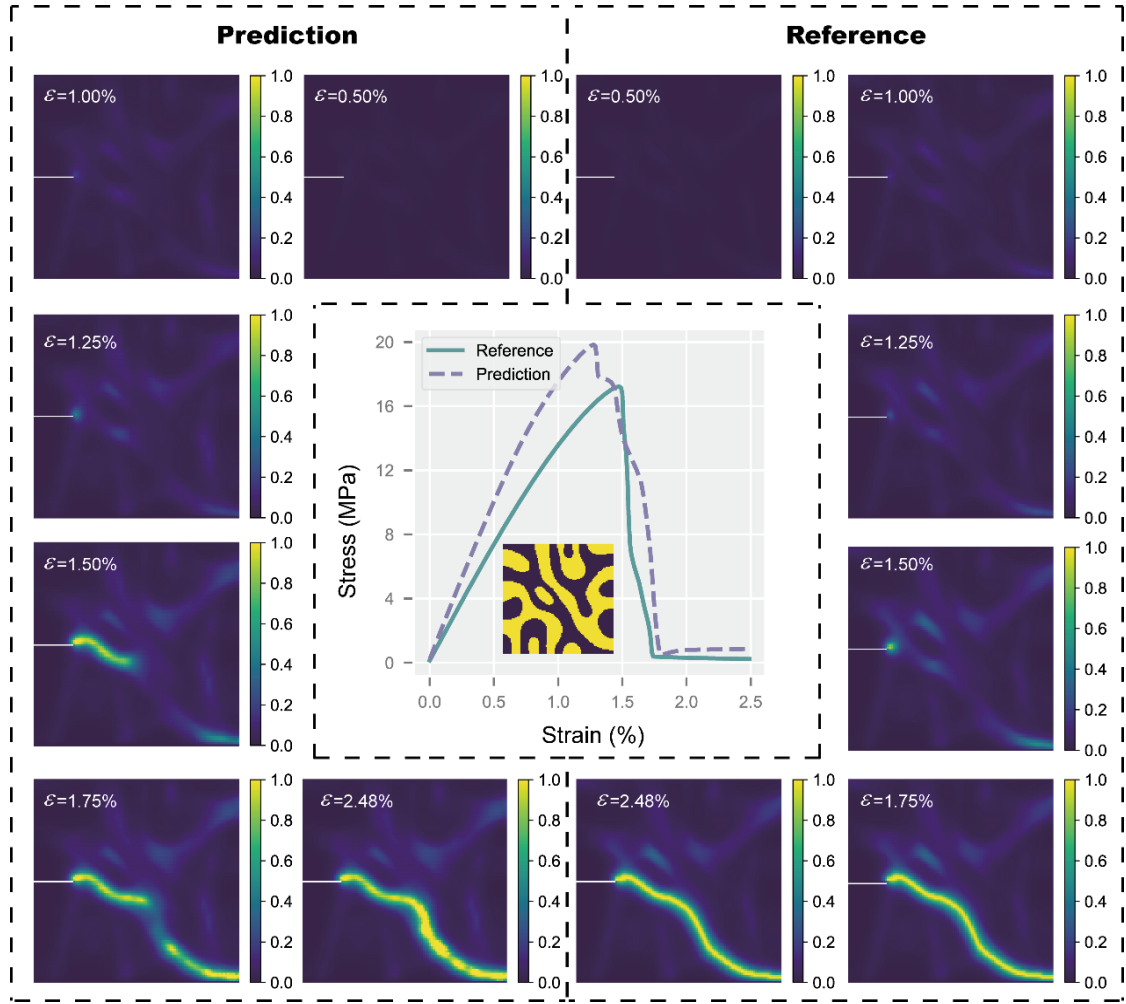

**Fig. S17. The out-of-sample long-prediction by Crack-Net and the reference from numerical simulation for the two-phase co-continuous structure.** The center of the figure is the predicted stress-strain curve and the reference, and the composite design is displayed at the bottom. In the composite design, the yellow region refers to the matrix, and the black region refers to the reinforcement phase. On the left side of the figure, predicted phase fields for various steps are displayed, while the right side exhibits the actual phase fields in corresponding steps obtained from numerical simulation. The white line in the phase field represents the initial condition.

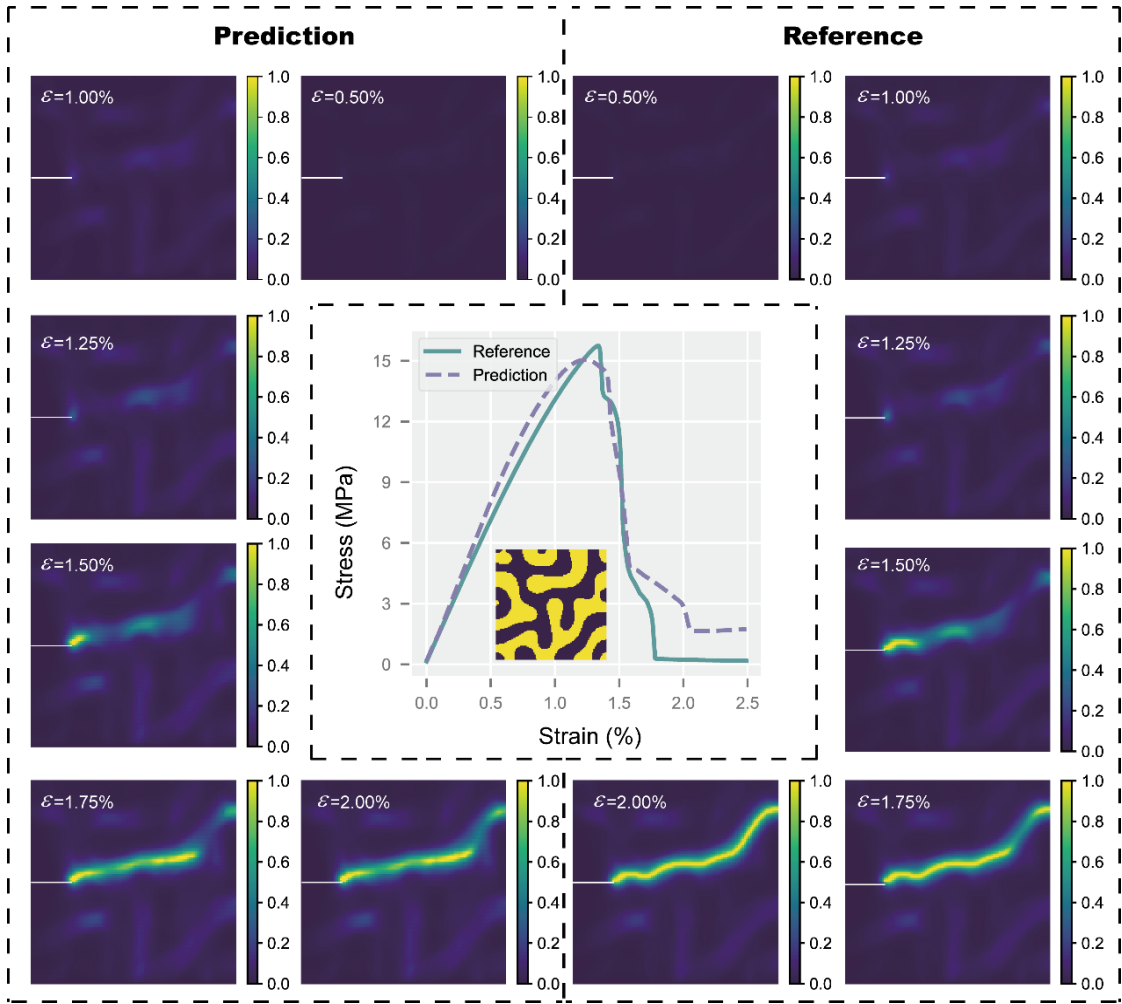

**Fig. S18. The out-of-sample long-prediction by Crack-Net and the reference from numerical simulation for the two-phase co-continuous structure.** The center of the figure is the predicted stress-strain curve and the reference, and the composite design is displayed at the bottom. In the composite design, the yellow region refers to the matrix, and the black region refers to the reinforcement phase. On the left side of the figure, predicted phase fields for various steps are displayed, while the right side exhibits the actual phase fields in corresponding steps obtained from numerical simulation. The white line in the phase field represents the initial condition.

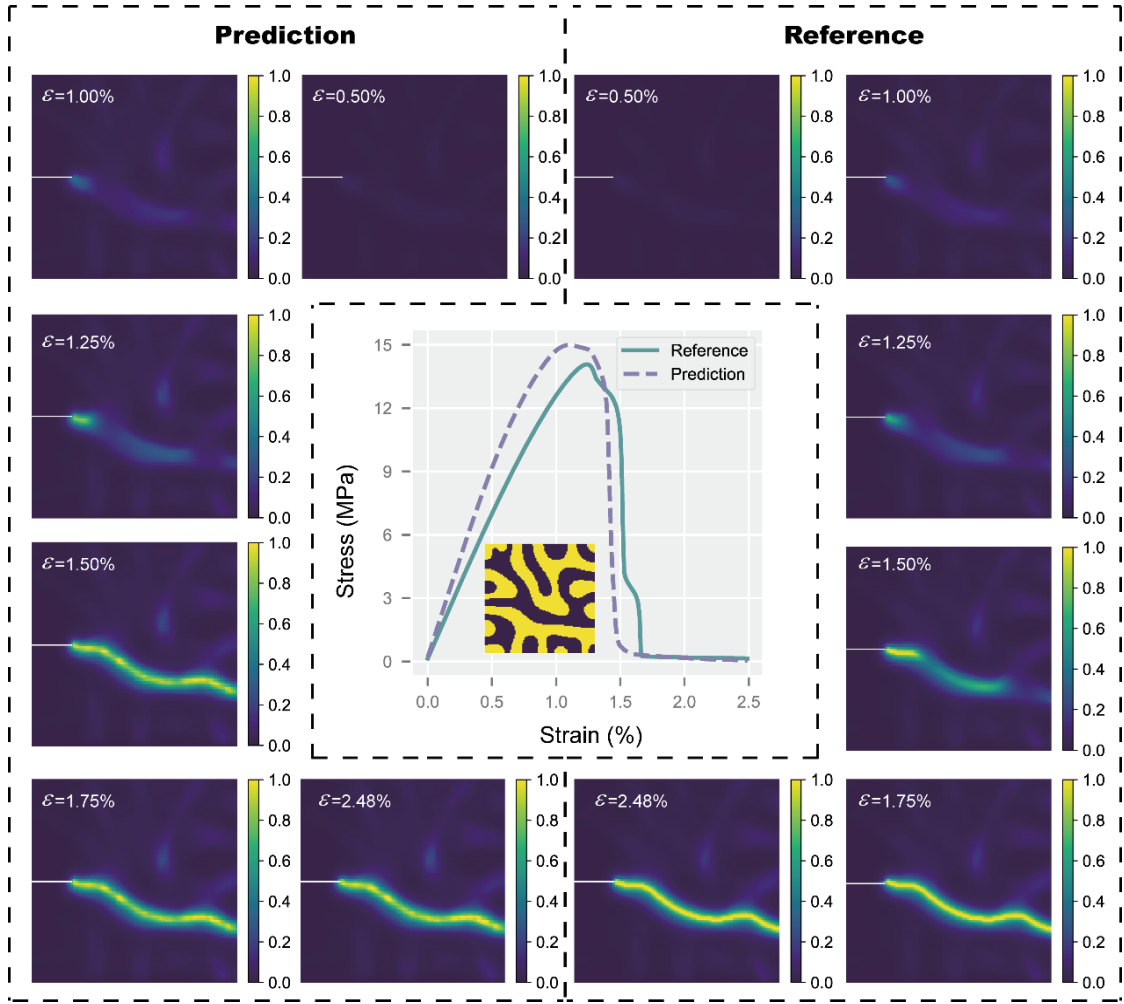

**Fig. S19. The out-of-sample long-prediction by Crack-Net and the reference from numerical simulation for the two-phase co-continuous structure.** The center of the figure is the predicted stress-strain curve and the reference, and the composite design is displayed at the bottom. In the composite design, the yellow region refers to the matrix, and the black region refers to the reinforcement phase. On the left side of the figure, predicted phase fields for various steps are displayed, while the right side exhibits the actual phase fields in corresponding steps obtained from numerical simulation. The white line in the phase field represents the initial condition.

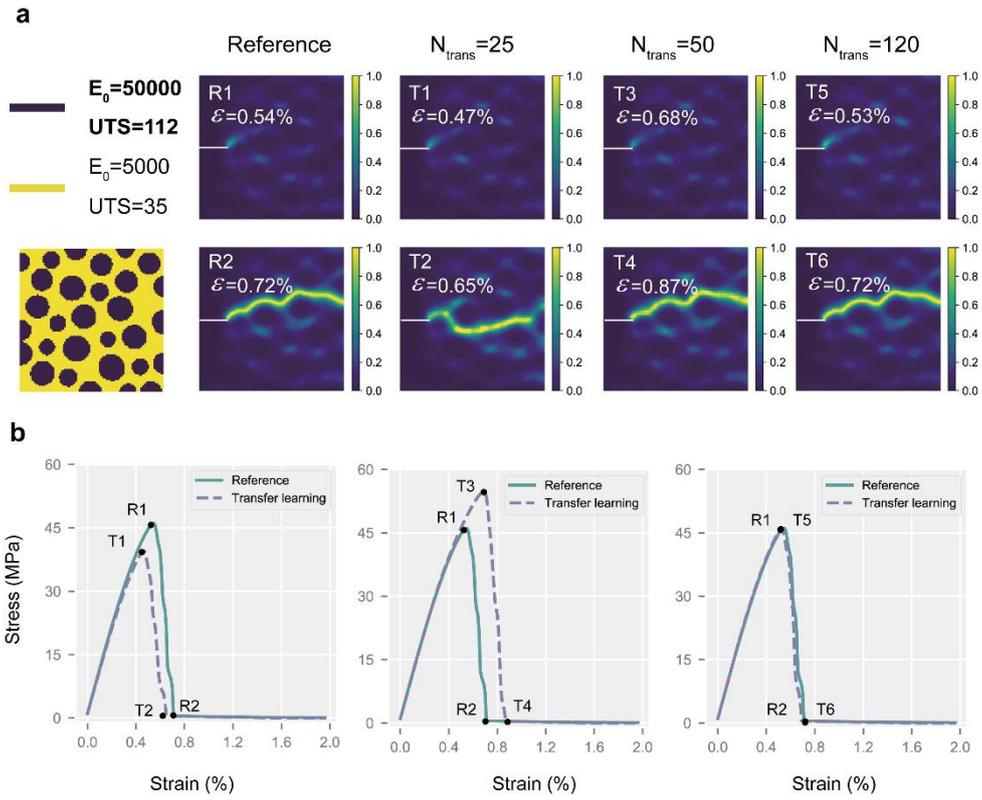

**Fig. S20. The performance of Crack-Net with different amounts of data for transfer learning.**
**(a)** The composite design and material attributes of the case (left), and the prediction and reference for two characteristic steps in the crack process by Crack-Net trained with different amounts of data for transfer learning, $N_{trans}$. **(b)** The predicted stress-strain curves by Crack-Net with different $N_{trans}$, and the corresponding reference. The characteristic steps are denoted on the stress-strain curves. The white line in the phase field represents the initial condition.

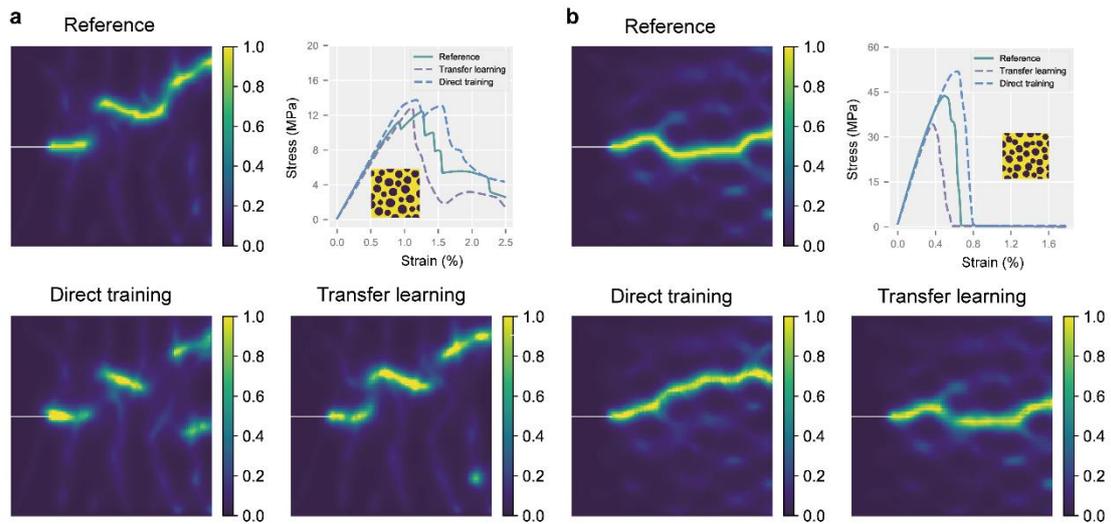

**Fig. S21. The comparison between direct training and transfer learning. (a)** The reference (upper left), the predicted ultimate crack phase field from Crack-Net directly trained by 50 cases (lower left), the predicted ultimate crack phase field from Crack-Net trained by transfer learning with 50 cases (lower right), and the comparison of stress-strain curves (upper right). In this case, the ultimate tensile strength (UTS) is 4.8, and the initial elastic modulus ($E_0$) is 100. **(b)** The reference (upper left), the predicted ultimate crack phase field from Crack-Net directly trained by 50 cases (lower left), the predicted ultimate crack phase field from Crack-Net trained by transfer learning with 50 cases (lower right), and the comparison of stress-strain curves (upper right). In this case, the ultimate tensile strength (UTS) is 112, and the initial elastic modulus ($E_0$) is 5000. The white line in the phase field represents the initial condition.

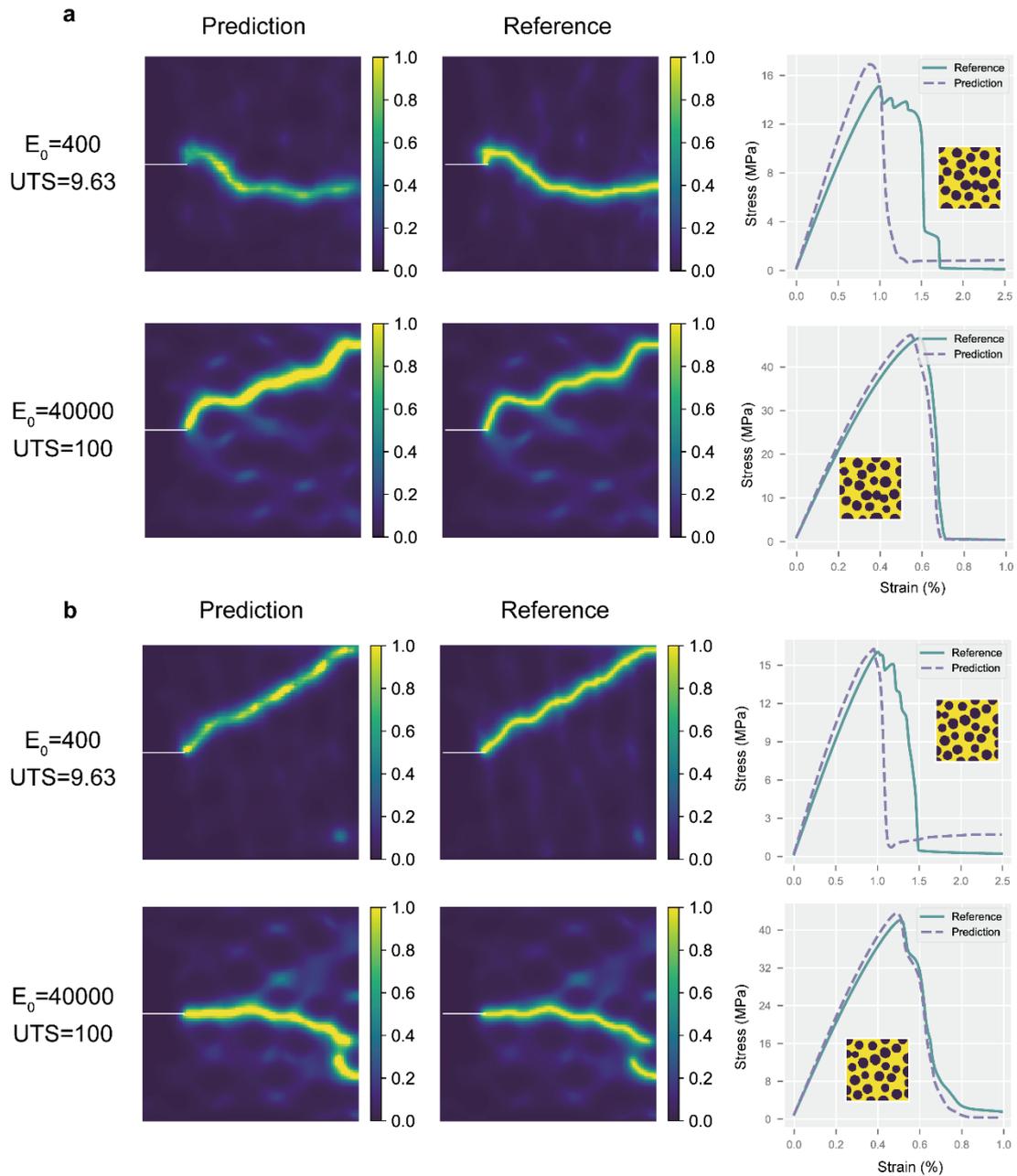

**Fig. S22. The predicted crack phase field and stress-strain curves for composites with the same design and different material properties. (a)** The prediction for composites with weak reinforcement phase (upper) and strong reinforcement phase (lower). **(b)** The prediction for composites with weak reinforcement phase (upper) and strong reinforcement phase (lower). The white line in the phase field represents the initial condition.

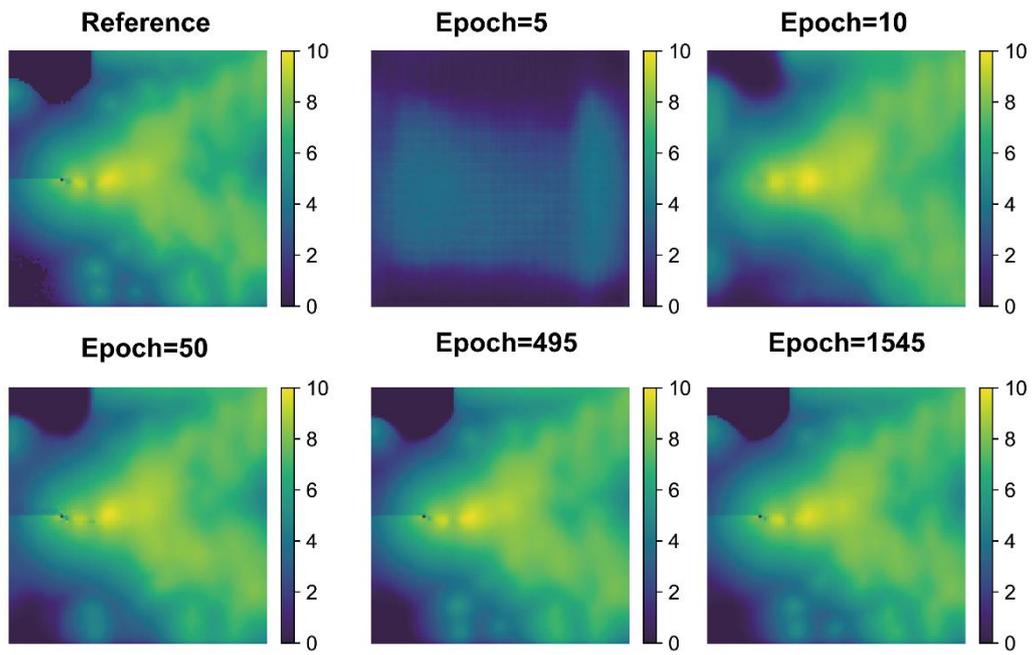

**Fig. S23. The predicted crack phase field in different epochs of the training process.** The epoch is the training epoch.

**Table S1. The simulation setup of the morphology in the dataset**

| Morphology | $c_{B0}$ | $E_B$ (GPa) | $V_{f,B}$ |
|---|---|---|---|
| PRC | 0.25 | 0.5 | ~55% |
| PRC | 0.25 | 0.5 | ~80% |
| PRC | 0.30 | 0.5 | ~54% |
| PRC | 0.30 | 0.5 | ~73% |
| PRC | 0.35 | 0.5 | ~57% |
| PRC | 0.35 | 0.5 | ~80% |
| PRC | 0.40 | 0.5 | ~65% |
| PRC | 0.40 | 0.5 | ~80% |

**Table S2. The simulation setup the morphology in the dataset for transfer learning**

| Morphology | $c_{B0}$ | $E_B$ (GPa) | $V_{f,B}$ |
|---|---|---|---|
| PRC | 0.25 | 0.1 | ~55% |
| PRC | 0.25 | 50 | ~55% |
| PRC | 0.30 | 0.1 | ~54% |
| PRC | 0.30 | 50 | ~54% |
| PRC | 0.35 | 0.1 | ~57% |
| PRC | 0.35 | 50 | ~57% |
| PRC | 0.40 | 0.1 | ~65% |
| PRC | 0.40 | 50 | ~65% |